\documentclass[pdflatex,sn-mathphys-num]{sn-jnl}

\usepackage{graphicx}%
\usepackage{multirow}%
\usepackage{amsmath,amssymb,amsfonts}%
\usepackage{amsthm}%
\usepackage{mathrsfs}%
\usepackage[title]{appendix}%
\usepackage{xcolor}%
\usepackage{textcomp}%
\usepackage{manyfoot}%
\usepackage{booktabs}%
\usepackage{algorithm}%
\usepackage{algorithmicx}%
\usepackage{algpseudocode}%
\usepackage{listings}%
\usepackage{subcaption}
\usepackage{comment}
\usepackage{url}

\theoremstyle{thmstyleone}%
\theoremstyle{thmstyletwo}%

\theoremstyle{thmstylethree}%

\begin{document}

\title[Article Title]{Towards a Domain-Specific Modelling Environment for Reinforcement Learning}

\author{\fnm{Natalie} \sur{Sinani}}\email{natalie.sinani@torontomu.ca}

\author{\fnm{Sahil} \sur{Salma}}\email{ssalma@torontomu.ca}

\author{\fnm{Paul} \sur{Boutot}}\email{pboutot@torontomu.ca}

\author*{\fnm{Sadaf} \sur{Mustafiz}}\email{sadaf.mustafiz@torontomu.ca}

\affil{\orgdiv{Department of Computer Science}, \orgname{Toronto Metropolitan University}, \orgaddress{\city{Toronto}, \state{ON}, \country{Canada}}}


\abstract{In recent years, machine learning technologies have gained immense popularity and are being used in a wide range of domains. However, due to the complexity associated with machine learning algorithms, it is a challenge to make it user-friendly, easy to understand and apply. Machine learning applications are especially challenging for users who do not have proficiency in this area.

In this paper, we use model-driven engineering (MDE) methods and tools for developing a domain-specific modelling environment to contribute towards providing a solution for this problem. We targeted reinforcement learning from the machine learning domain, and evaluated the proposed language, reinforcement learning modelling language (RLML), with multiple applications. 
The tool supports syntax-directed editing, constraint checking, and automatic generation of code from RLML models. The environment also provides support for comparing results generated with multiple RL algorithms. 
With our proposed MDE approach, we were able to help in abstracting reinforcement learning technologies and improve the learning curve for RL users.
}


\maketitle

\section{Introduction}
\label{sec:Introduction}

The advent of artificial intelligence and machine learning technologies marks a significant shift in the landscape of software and systems, with groundbreaking developments continually emerging. Among these, reinforcement learning (RL)~\cite{Brunton2019}, a fundamental paradigm of machine learning, is gaining prominence. Originally notable in gaming, where machines could outperform expert human players, RL is now increasingly relevant in dynamic and adaptive environments with applications spanning from healthcare, finance to self-driving cars. However, the complexity of RL algorithms poses a significant barrier, often necessitating ML domain expertise for effective implementation and use. Despite the growing demand, current professional profiles lack the extensive skills sets necessary for its adoption~\cite{Bucchiarone2020}.

Just as the need for intelligence in various application areas has led to the integration of machine learning algorithms into more user-friendly interfaces, there is a similar demand in the realm of RL. Data scientists typically rely on specific libraries such as OpenAI Gym, TensorFlow Agents (TF-Agents), and Stable Baselines for implementing RL algorithms, which requires a deep understanding of the intricate interfaces. Additionally, RLlib, as part of the Ray Project, is gaining popularity due to its scalability and support for multi-agent RL. This technical complexity underscores the necessity for more accessible, abstracted systems that can democratize the use of RL for non-expert users~\cite{Baier2019}. However, majority of the research work in the RL domain focuses on enhancing algorithms and approaches, and in achieving better accuracy and results in prediction and learning, but there is not much work done in the area of simplifying the application of these complex algorithms by offering more convenient and user-friendly techniques and tools for non-technical users.

Model-driven engineering (MDE) can contribute to this challenge by providing enablers to directly express and manipulate domain-specific problems~\cite{Bucchiarone2020}. Domain-specific languages (DSL) in MDE aim to reduce complexity with the use of abstraction.  
In response to this challenge, we propose a DSL tailored for RL, named Reinforcement Learning Modelling Language (RLML). This initiative mirrors the approach in the broader machine learning field, where abstraction and user-friendly interfaces are increasingly recognized as essential. RLML focuses on model-free algorithms, which are more widely used and extensively tested, to ensure broad applicability. RLML was developed on the JetBrains MPS platform~\cite{Voelter2013}, providing an integrated modelling environment that streamlines the creation, execution, and analysis of RL models. Furthermore, to bridge the gap between Python, the predominant language in ML, and Java, we have implemented model-to-code transformations for both languages, enhancing the accessibility of RL algorithms.

This paper is organized as follows: Section~\ref{sec:background} provides a brief background on reinforcement learning. Section~\ref{sec:language} presents our domain-specific modelling language, RLML. Section~\ref{sec:environment} covers the domain-specific modelling environment built using the language workbench, MPS JetBrains. Section~\ref{sec:validation} provides a description of the reinforcement learning applications we used to validate RLML. Section~\ref{sec:relatedwork} discusses related work and Section~\ref{sec:conclusion} concludes the paper.

\section{Background}
\label{sec:background}

This section provides necessary background on model-driven engineering and reinforcement learning.

\subsection{Model-Driven Engineering}

Model-Driven Engineering (MDE) is an approach to software development where models take a central role \cite{Schmidt2006}. In MDE, abstract representations of the knowledge and activities of a domain are created, which form the basis for code generation, analysis, and simulation. This approach emphasizes the separation of concerns, allowing developers to focus on domain-specific logic rather than low-level implementation details. MDE is particularly effective in managing complexity and enhancing productivity, as it allows for high-level abstraction and automation in the software development process.

Models in MDE have to conform to some modelling language, which may be graphical or textual. These languages have three main ingredients: abstract syntax , concrete syntax, and semantics. In addition to general-purpose modelling languages (e.g., UML), domain-specific languages (DSL) can also be developed, which are tailored to specific problem domains and simplify complex concepts, making models more accessible and intuitive for users.
Expert knowledge is used to define DSLs that work at a higher level of abstraction than general-purpose languages. This allows a tight fit with the domain and changes in the domain can be easily incorporated into the language. The use of such languages separate the domain-experts work from the software developers work and also make it easier for non-technical people to learn and avoid errors in the initial development phases. They also provide support for text or code generation, which automates the process of creating executable source code and/or textual artifacts (e.g., documentation, task lists, etc.) from the domain-specific models. The developers are not required to manually write and maintain these artifacts, which significantly improves productivity and reduces defects in the implementation, thus improving software quality.

\subsection{Reinforcement Learning} 

Reinforcement learning (RL) is an area of machine learning concerned with how intelligent agents ought to take actions in an environment in order to maximize the notion of rewards. It is a self-teaching system trying to find an appropriate action model that would maximize an agent’s total cumulative reward, by following the trial and error method. In general, the RL algorithms reward the agent for taking desired actions in the environment, and punishes i.e., grants negative or zero rewards, for the undesired ones \cite{Brunton2019}. The following are the key components that describe RL problems.


\begin{itemize}
\item{Environment}: The RL environment~\cite{Graesser2019} represents all the existing states that the agent can enter. It produces information that describe the states of the system. The agent interacts with the environment by observing the state space and taking an action based on the observation.
Each action receives a positive or negative reward, which informs the agent on selecting the next state.
\item{Agent}: This is represented by an intelligent reinforcement learning algorithm that learns and makes decisions to maximize the future rewards while moving within the environment. 
\item{State}: The state represents the current situation of the agent. It should be noted that term \emph{state} in RL is different from the meaning of \emph{state} in MDE in the context of state machines~\cite{harel1987statecharts}. 
\item{Action}: The mechanism by which the agent transitions between states of the environment.
\item{Reward}: The environment feeds the agent with rewards. Rewards are numerical values that the agent tries to maximize over time. They are received on each action and may be positive or negative.
    
\end{itemize}

Reinforcement learning algorithms estimate how \emph{good} it is for the agent to be in a certain state. This estimation is the calculation of what is known as a value function \cite{Montague1999}. The value function gets measured based on the expected future rewards that the agent will receive starting from a given state \emph{s}, and according to the actions that the agent will make, and this is referred to as the \emph{expected return}.
The goal of a reinforcement learning algorithm is to find the optimal policy for an agent to follow that maximizes the expected return. An optimal policy will have the highest possible value in every state. The optimal policy is implicit and can be derived directly from the optimal value function. There are many different approaches to find the optimal policy. They are mainly categorized as model-based or model-free learning algorithms, in addition to deep reinforcement learning. It is worth mentioning that this categorization is not comprehensive and it is often blurry~\cite{Brunton2019}.
            
When the model of the environment is available, which is the case with model-based algorithms, the RL problem is simpler and we can utilize policy iteration or value iteration algorithms. To learn the optimal policy or value function we either have access to the model (environment) and we know the probability distribution over states, or we try to build a model. When the agent knows its model and probability distribution over states it can use it to plan its next moves. However, it is more challenging when we are dealing with model-free algorithms, and it is often the case in real life scenarios, where the agent does not know the environment and needs to discover it.
As stated by Sutton and Barto: \emph{model-based methods rely on planning as their primary component, while model-free methods primarily rely on learning} \cite{Montague1999}. Finally, deep reinforcement learning incorporates deep learning techniques and algorithms in order to learn the model \cite{silver2015}.  
            
            \begin{figure}[!tbh]
               \centering
                \includegraphics[width=0.85\columnwidth]{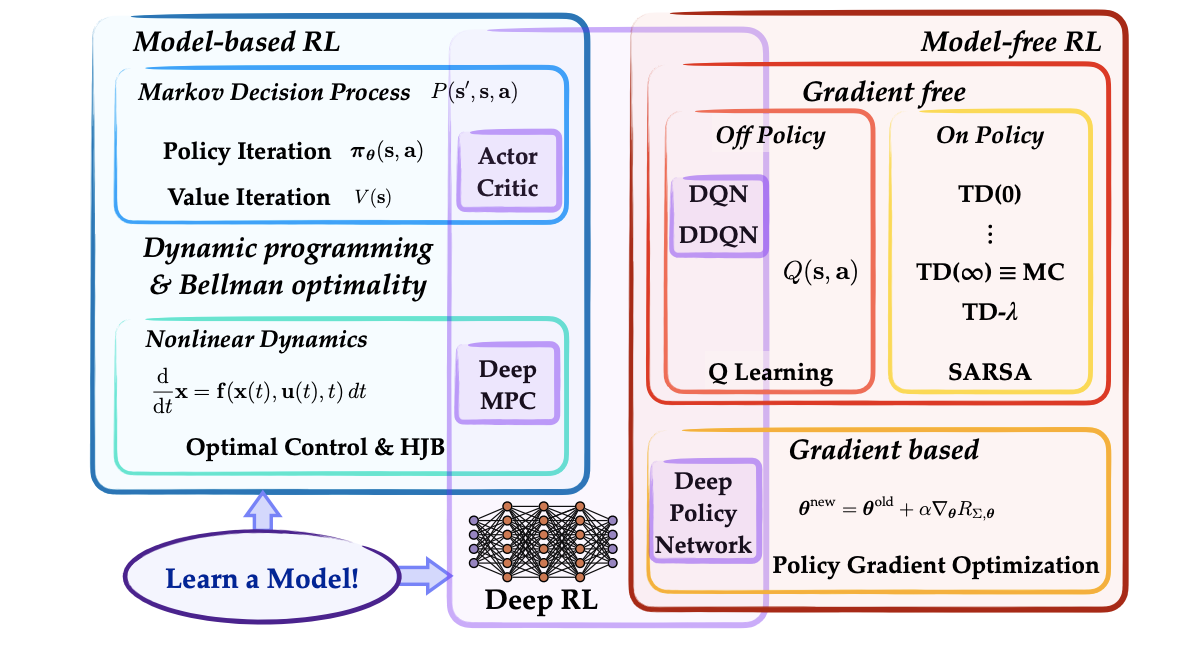}
                \caption{Reinforcement Learning (RL) Classification (Adapted from \cite{Brunton2019})}
                \label{fig:RLCats}
            \end{figure}

In model-free approaches, the agent learns and evaluates how good actions are by trial and error method. The agent relies on the past experiences to derive the optimal policy (described earlier). 
Various available algorithms in this approach include (see Fig. \ref{fig:RLCats}), Monte Carlo learning, Model-Free Actor Critic, SARSA: State–action–reward–state–action learning, DQN and Q-Learning.
            
In this work, we have focused on model-free gradient-free algorithms. The objective of such algorithms is to maximize an arbitrary score, which is the value function, hence also referred to as value-based algorithms. In value-based methods, the algorithm does not store any explicit policy, only a value function. Some algorithms, such as actor critic, are both value and policy-based.
Q-Learning, perhaps, is one of the most dominant model-free algorithm which learns the Q-function directly from experience, without requiring access to a model.

Since we are using both model-driven engineering and machine learning technologies in our research, we would like to clarify that models have different meaning in each of these two fields. Models in MDE refer to software models and are an abstract representation of the elements that define the software and system domain~\cite{Schmidt2006}. On the other hand, models in machine learning are algorithms that contain defined instruction and mathematical formulations \cite{Jiang2021}. Models in ML can be trained to recognize certain patterns in provided data.

\section{Modelling Language Design}
\label{sec:language}

In this section, we describe the proposed DSL for reinforcement learning, RLML. The core language concepts was designed based on the main elements representing the reinforcement learning problem and solution algorithms. This includes the environment representing the reinforcement learning problem, and the agent that is learning and exploring in the environment. 
        
\subsection {RLML Abstract Syntax}

\begin{figure}[!tbh]
            \centering
            \includegraphics[width=1\textwidth]{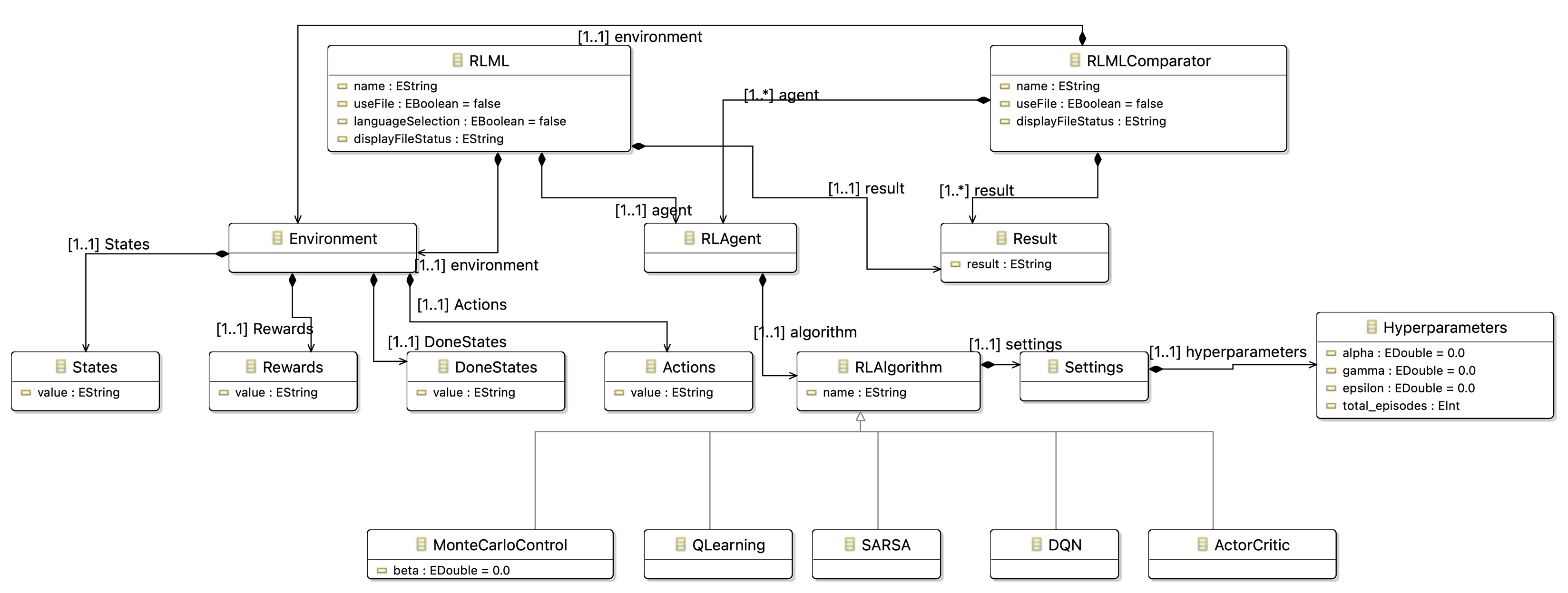}
            \caption{RLML Metamodel}
            \label{fig:RLMLMetamodel}
        \end{figure}

Reflecting the RL domain concepts, RLML mainly consists of an \emph{environment} element and an \emph{agent} element, and additionally, the \emph{result} element. Successively these elements contain all the other details involved in solving an RL problem. 
Similarly, the \emph{RLMLComparator} consists of the same elements as RLML, except it can have multiple agent elements as well as corresponding number of result elements.
Fig. \ref{fig:RLMLMetamodel} presents the metamodel for RLML.
        
\begin{itemize}
\item{RLML:} The RLML element is the root element of all the other elements in the language, 
and contains the environment element, RL agent and the result. It also has a string property which represents the name of the project, along with other properties that lets the user decide an input method and a run language method.

\item{RLML Comparator:} This is another root element which is almost a replica of RLML and contains all the other elements in the language except RLML. 
It contains the environment element, multiple RL agent elements, multiple result elements and a string property which represents the name of the project.
        
\item{Environment:} This represents the RL problem environment, which is used to describe the RL problem and the goal that the agent needs to reach. It is broken down into states, actions, terminal states and rewards elements. Each one of these elements contains a value property, which expects a string value and have associated constraints. It should be noted that the values for the environment can also be supplied through a file containing values that adhere to necessary constraints for all elements.
        
\item{Reinforcement Learning Agent:} The RL agent in the domain is represented by the RL algorithm, which will be used to to solve the RL problem, given by the RL environment. 
        
\item{Reinforcement Learning Algorithm:} 
This is specialized into the many different RL algorithms which can be chosen and implemented to solve the RL problem. It holds the settings property with reference to the required settings and parameters to tune an RL algorithm. All the child RL algorithms will inherit the settings property. 
The settings element carries the common RL algorithm parameters. A specific type of RL algorithm can have its own specific properties.

The algorithms currently covered in the language include Q-Learning, SARSA, Monte Carlo, Actor Critic and DQN all of which fall under model-free RL. The metamodel is easily extensible to support addition of more RL algorithms as the language matures, to cover more tests cases and broader RL problems.
        
\item{Settings and Hyperparameters:} The settings element contains hyperparameters, which include all the common properties for the selected RL algorithms. The hyperparameters element contains alpha (the learning rate), gamma (the discount factor), epsilon (specifies the exploration rate), and total episodes (total number of episodes to train the agent). More parameters can be added for specific algorithms in their individual concepts
        
\item{Result:} The result element contains a result string property, and is used to display the results of running the chosen algorithm. 

\end{itemize}

\subsection {RLML Concrete Syntax}
\label{sec:RLML concrete syntax}

One of the goals of RLML is to reduce the complexity involved in implementing RL applications. 
RLML uses textual concrete syntax and can be modelled as a simple configuration-like properties file, as shown in the sample model in Fig.~\ref{fig:RLMLModelMPS}.
        
The inspiration of the RLML concrete syntax comes from YAML representation, which is a human-readable data format used for data serialization. It is used for reading and writing data independent of a specific programming language. 
Another significant aspect for this concrete syntax is its relevance in model-free algorithms, where the a dynamic state-action space is required for the agent's actions. This interaction-focused approach is key in model-free reinforcement learning, allowing the agent to effectively learn and refine its strategy through direct experience, even in complex and variable scenarios.
As per the abstract syntax, the model needs to specify the project's name, the environment element properties (the states, actions, rewards and terminal states), and the agent's RL algorithm type and the settings for that algorithm.

\subsection {RLML Constraints}
\label{subsec:rlml-constraints}

The property values are considered valid when they are in a format that RLML can use to implement the chosen RL algorithm. To ensure that the user is entering valid properties, we defined the following DSL validation constraints.

\begin{itemize}
\item{States Constraint:} States property value is expecting a string representation of all the possible states an agent can move within the current world or the environment of the current task. The value of the states property must be a comma-separated list of state strings, within square brackets. The individual state names cannot have comma or spaces.\newline
Valid example: [A, B, C, D, E, F]
        
\item{Actions Constraint:} The possible actions that the agent can take for each state of the states array. This value is also in string format and expects a two-dimensional array of indexes. The array of indexes contain the index values of the states that the agent can go to, starting from the given state. Each array is a comma-separated list within square brackets. The constraint validates the format of the provided string value and checks that the length of actions array element is equivalent to the length of the states array. In the valid example below, we can see that there are six arrays of indexes to match the length of the example array for states.\newline
States example: [A, B, C, D, E, F]\newline
Valid Example: [[1,3], [0,2,4], [2], [0,4], [1,3,5], [2,4]]
        
\item{Rewards Constraint:} The rewards property value is similar to the actions property value. 
In this case, the two-dimensional array contains reward values the agent will receive when moving from the given state to other states in the environment. The RL algorithm will eventually learn to move towards the states that give maximum future rewards and ignore the ones that do not give rewards. Each array is a comma-separated list within square brackets. Similar to actions value validation, the rewards constraint validates the format of the string and checks that the length of the rewards array is equal to the length of states array and the length of individual rewards elements, is equivalent to the length of the states array. In the valid example below, there are also six arrays of six reward values to match the length of states example array.\newline
States example: [A, B, C, D, E, F] \newline
Valid Example: [[0,0,0,0,0,0], [0,0,100,0,0,0], [0,0,0,0,0,0], [0,0,0,0,0,0], [0,0,0,0,0,0], [0,0,100,0,0,0]]
        
\item{Terminal States Constraint:} In the RL domain, the terminal states is a subgroup of all the states that can end a training episode, either because it is the goal state or because it is a terminating state. Therefore, the terminal states array should provide a smaller string array than the states array. The terminal states constraint ensures the format of the string value provided is a comma-separated list within square brackets and checks that this array is a sub-array of the states example array.\newline
States example: [A, B, C, D, E, F]\newline
Valid Example: [C]
\end{itemize}
\section{Domain-Specific Modelling Environment}
\label{sec:environment}

This section describes the tool support developed for RLML.

\subsection{RLML Features}

A modelling environment has been designed and developed to create RLML models. Translational semantics have been implemented to support execution of the models through the environment. 
The modelling environment supports use of different agents as well as displays the output of the RL training in the environment. 

In our project, which supports both Java and Python, we focus on maintaining algorithmic uniformity. Thus we have the same algorithms for both general purpose languages. Concurrently, efforts were made to enhance Java's RL capabilities, ensuring it remains a viable option for those preferring or requiring it. Our balanced approach enhances the project's overall utility, catering to the diverse needs of the RL community and maintaining inclusivity across programming preferences.

We have also incorporated a capability for users to save their trained RL models. This addition serves to facilitate the retention and subsequent utilization of these models, offering researchers a valuable resource. Additionally, we developed support for running multiple algorithms simultaneously, presenting data for each distinct variation. This functionality not only allows users to compare and analyze different algorithmic approaches side by side but also facilitates a deeper understanding of how variations in parameters affect outcomes (see Fig.~\ref{fig:RLMLComparator}). It provides a robust platform for experimentation, enabling users to efficiently identify the most effective algorithms and parameter settings for their specific use cases. This multi-algorithm capability greatly enhances the tool's utility in complex scenarios, making it an invaluable asset for both research and practical applications in diverse fields where nuanced algorithmic comparisons are essential.

Recognizing the complexity of RL inputs and the impracticality of manual entry in some cases, we have enhanced RLML with the capability to import values through a text file. This feature allows users to select a file (see Fig.~\ref{fig:RLMLTextFile}), which is then processed to ensure it contains valid data. Upon confirmation of valid input, the system automatically populates the \emph{States}, \emph{Actions}, \emph{Rewards}, and \emph{Done States}. This addition significantly enhances the versatility of the tool, making it suitable for use cases that involve large input sizes.
    
The proposed DSL for the reinforcement learning domain is developed using MPS (Meta Programming System), which is a modelling language workbench developed by JetBrains~\cite{MPS}.

    \begin{figure}[!tbh]
            \centering
            \begin{minipage}[t]{0.585\textwidth}
            \centering
            \includegraphics[width=1.0\linewidth]{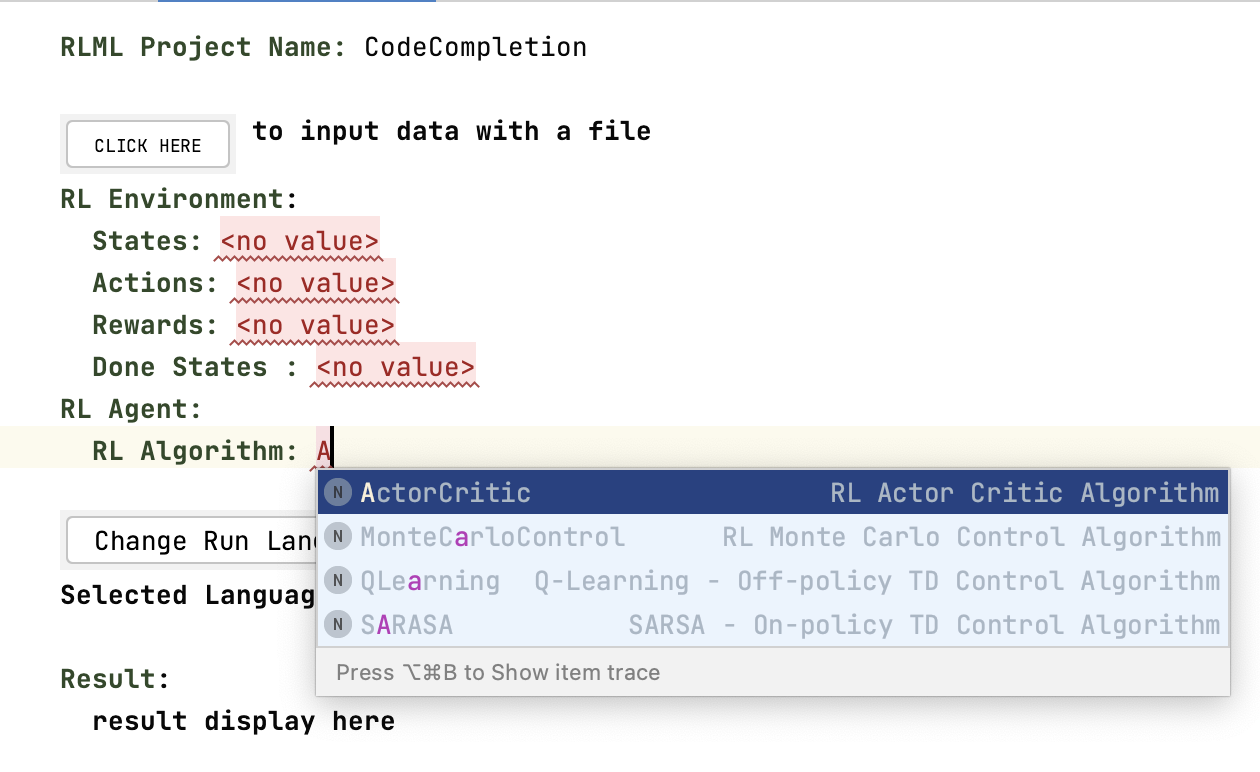}
            \caption{RLML Environment: Code Completion}
            \label{fig:MPSCodeCompletion}
            \end{minipage}
            \begin{minipage}[t]{0.4\textwidth}
            \centering
            \includegraphics[width=1.0\linewidth]{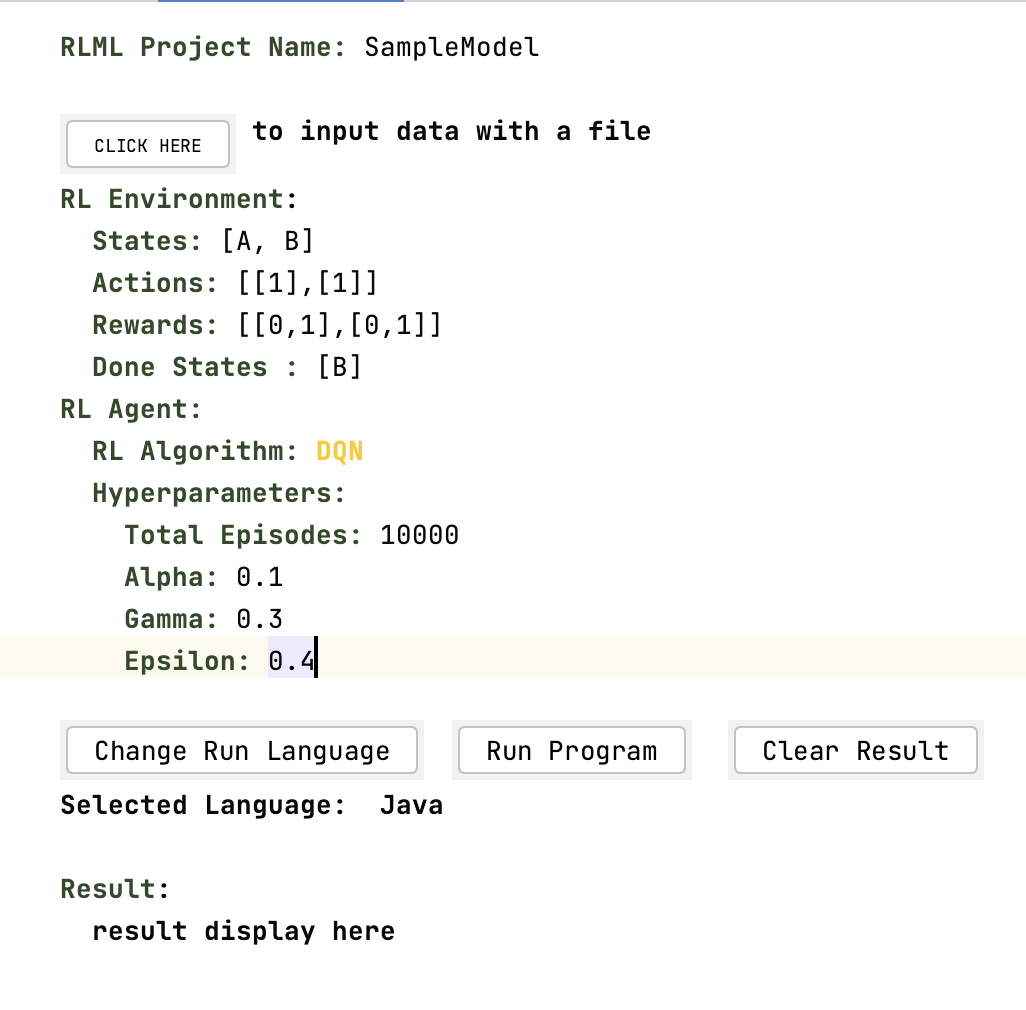}
            \caption{Sample RLML Model}
            \label{fig:RLMLModelMPS}
            \end{minipage}
    \end{figure}

\subsection{RLML Editor}
\label{sec:MPS Environment}
MPS is a language workbench which provides a tool or set of tools to support language definition, and it implements language-oriented programming. MPS is an integrated development environment (IDE) for DSL development, which promotes reusability and extensibility.
The language definition in MPS consists of several aspects: structure, editor, actions, constraints, behaviour, type system, intentions, plugin and data flow. Only the \emph{structure} aspect is essential for language definition and the rest are for additional features. These aspects describe the different facets of a language. 
        
We have employed the structure, editor and constraints aspects in the RLML definition. The structure aspect, defines the nodes of the language Abstract Syntax Tree (AST), known as concepts in MPS. The editor aspect describes how a language is presented and edited in the editor and it enables the language designer to create a user interface for editing their concepts. Constraints describe the restrictions on the AST.

\begin{itemize}
\item{\emph{RLML structure:}} The structure aspect contains the concepts that represent the RLML metamodel. Each concept consist of properties and children, reflecting the properties and the relationships in the RLML metamodel, shown in Fig. \ref{fig:RLMLMetamodel}.

\item{\emph{RLML editor:}} 
The RLML editor aspect is configured to define RLML's concrete syntax as described and illustrated earlier in Fig. \ref{fig:RLMLModelMPS}). The concept editor for RLML root element contains ``Click Here", ``Browse File", ``Change Run Language", ``Run Program" and ``Clear Result" buttons (see Fig. \ref{fig:RLMLModelMPS}). The first button (Click Here) refers to input via file, clicking the button will show or hide the ``Browse File" button. The second button (Browse File) opens a choose file wizard (see Fig. \ref{fig:RLMLTextFile}). Next button (Change Run Language) switches between running the python vs java code. The next button (Run Program) gets RLML model's generated code based on the previous selection, runs the program in the MPS environment and displays the result in the designated text area of the editor. The ``Clear Result" button clears the result. This interface simplifies the environment setup and allows for obtaining generated code in any supported language. Additionally, it enables the execution of this code and the display of results within the RLML environment..
        
With the support for automatic code completion in MPS, the environment shows suggestions as the user creates the RLML model. Code completion feature helps the RLML user to see the list of available reinforcement learning algorithms (refer to Fig. \ref{fig:MPSCodeCompletion}) and choose the one which can solve the targeted RL problem.

\begin{figure*}[!tbh]
        \centering
        \includegraphics[width=0.85\linewidth]{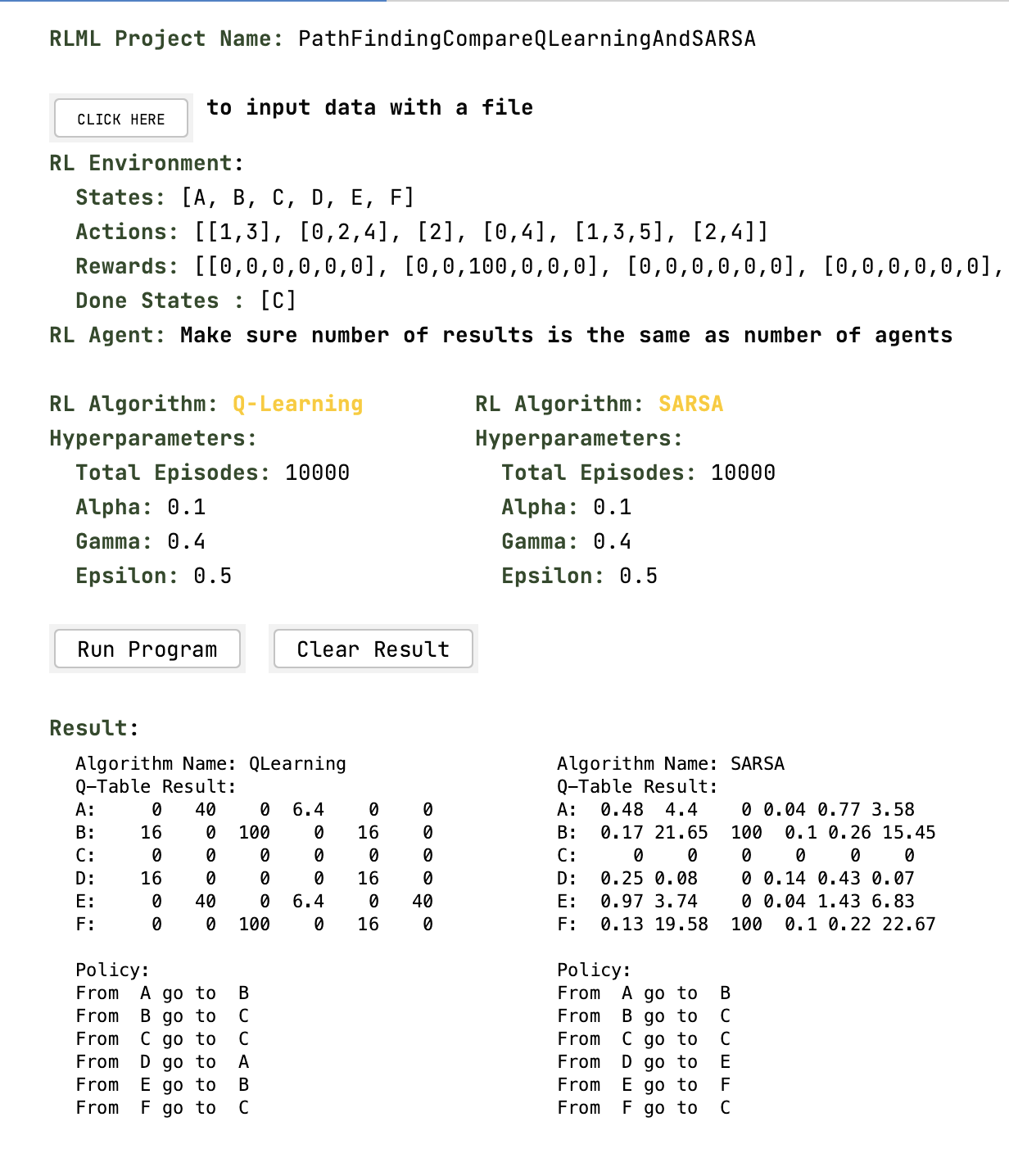}
        \caption{RLML Environment: Comparator}
        \label{fig:RLMLComparator}
            
\end{figure*}
    
\begin{figure}[!tbh]
        \centering
        \includegraphics[width=1\linewidth]{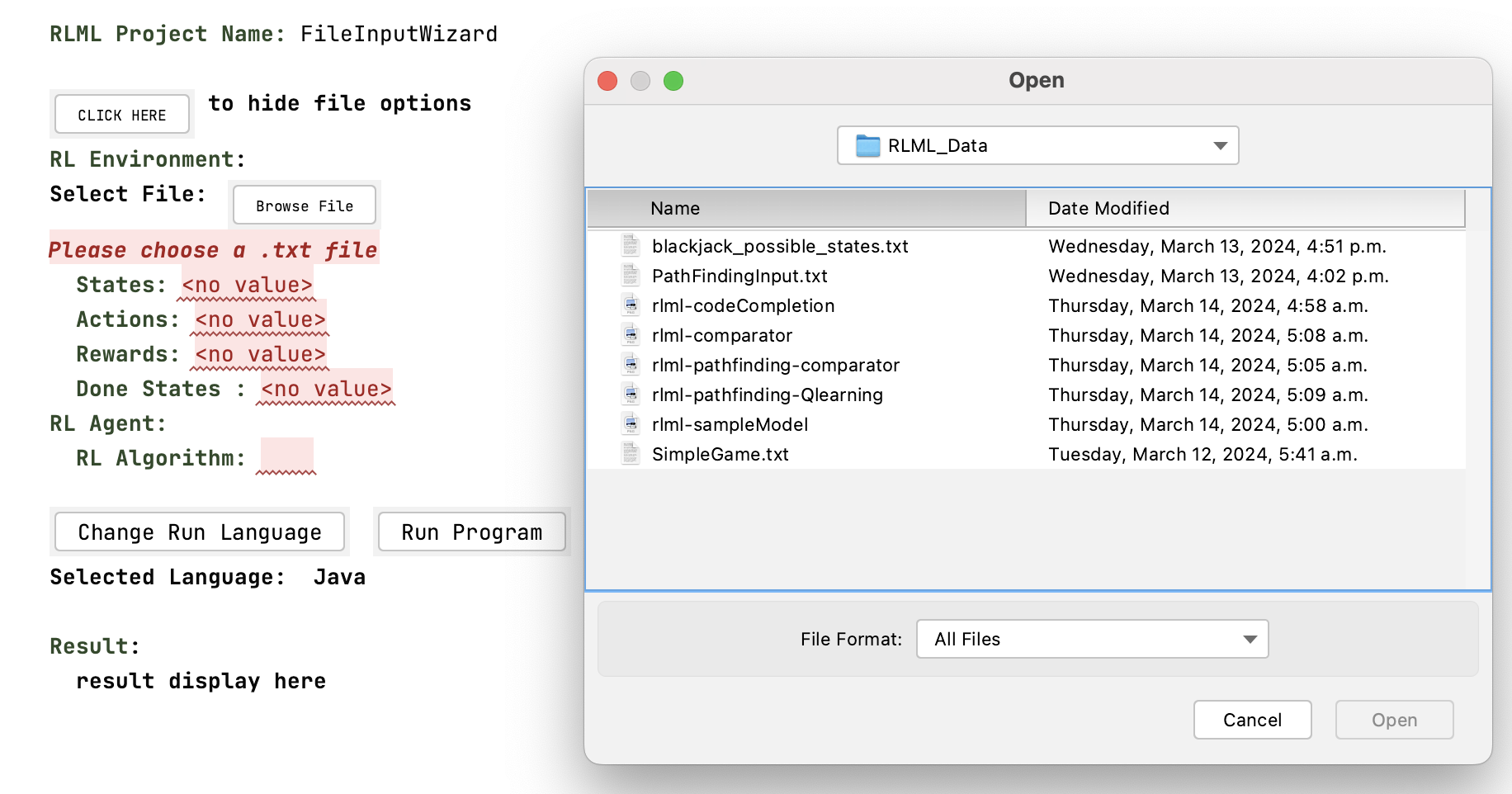}
        \caption{RLML Environment: Text File Input Wizard}
        \label{fig:RLMLTextFile}      
\end{figure}

\item{\emph{RLML constraints:}} The validation constraints are implemented using MPS's concept constraints capability. For each defined structure concept, we can develop a concept constraint to validate it. RLML constraints aspect reflect RLML's constraints (explained in Sec.~\ref{subsec:rlml-constraints}), which are the actions, rewards, states and terminal states constraints.
The sandbox solution in MPS facilitates implementing the developed language and holds the end user code. Fig.~\ref{fig:RLMLModelMPS} shows an example of an RLML model in MPS.

\end{itemize}

\subsection{RLML Code Generation}
This work aims to provide abstractions to reduce the complexity associated with reinforcement learning problems and algorithms by generating runnable code from the RLML models. Generators define possible transformations between a source modelling language and a target language, typically a general purpose language, like Java or Python. 
        
For our proposed language, we implemented the model to code transformation to generate code from RLML models.  
We have used a root mapping rule and reduction rules for our code generation.
While Java is directly supported by MPS, it is limited in generating code for Python. Since we want to generate python code, we utilize an open-source MPS module\footnote{https://github.com/juliuscanute/python-mps}. This module allows us to extend MPS's capabilities to generate Python, applying similar model-to-text transformations as with Java. This integration enhances the versatility of our tool, supporting a wider range of programming languages and accommodating a broader user base.

\begin{itemize}
\item{Root mapping rule:} RLML's generator module contains two root mapping rules, one for the RLML element and other for RLMLComparator element, which are the root elements of RLML. The rule specifies the template to transform the RLML element or RLML concept in MPS, into a valid general purpose language class with fields and methods corresponding to those in RLML element's properties and children.

\item{Reduction rule:} The generator module contains reduction rules for all supported algorithms. 
Supporting more RL algorithms simply means extending the language with additional RL algorithm concepts and their reduction rules/transformation rules. However, it is important to note that since Python does not have native support in MPS, the reduction rules cannot be used. To extend an algorithm in Python, we need to add a new function definition to the \texttt{mapRLMLmain.py} file. 
As part of our contributions, we are building a Java library of RL algorithms, to be used with MPS. While Python implementations are typical for RL, developers and students with Java expertise will find such a library quite beneficial. 
\end{itemize} 
        
Using these transformation rules, MPS can transform an RLML model to runnable code. The generated file contains more than 1500 lines of code for Java and about 300 lines for Python. This emphasizes the simplicity offered by RLML. The name of the file will be mapped to RLML element's project name property, and it contains a method called \texttt{run} which implements the RL algorithm calculations based on the reduction rules for Java or the function definition for Python defined earlier. In the file, the reduction rules part is the result of the \emph{reduce} transformation rule implementation. 

In user experience for the language, we have integrated UI elements like buttons into the MPS Editor and enabled file importation with data validation against set constraints. Addressing MPS's inability to support Python, we devised innovative solutions for Python code generation and execution within MPS. These enhancements not only streamline the RLML user experience but also broadly benefit MPS's community, particularly in modelling language development.

A video demonstrating the RLML modelling environment is available at \url{https://cs.torontomu.ca/~sml/demos/rlml.html}.

\subsection{Discussion}
\label{section:challenges}

Most machine learning libraries are widely available as Python libraries and not as Java libraries, hence it was challenging to find Java libraries to support RL algorithms. For the few available libraries, they were not fully supported by MPS. We were able to overcome this challenge by implementing algorithms in Java manually and using an open source MPS module to support Python code generation. 

A second issue was that RL problems do not have fixed input format from actions, rewards and states perspective. Therefore, it was not straightforward to come up with a format for those inputs. More validation and mapping is needed for broader RL problem coverage and code generation. This could be fixed by using more model-based algorithms where the agent is given an environment it can interact with to solve the problem.

Addressing the challenge of handling large data set inputs, our tool offers a feature for data input through a file. This method is particularly useful for adding large inputs efficiently. However, it is important to note that this feature requires the text files to adhere to a specific format (as defined in the concrete syntax of the language). Consequently, users need to either manually create these files or generate them specifically for use with this tool, considering the format requirements. In our tests, we were able to use a large language model (LLM)~\cite{radford2019-openai} along with precise prompts to generate these files.

We added support to save the RL model which can be useful for researchers to reuse the trained model. However, the limitation of this \emph{save} feature is it cannot be reused if the original parameters are changed. Effectively, reusing the saved model would be the same as increasing the number of episodes to train.

In practical applications of reinforcement learning, selecting appropriate states and actions can be quite complex, especially in real-world scenarios. Recognizing this, our future work on RLML will focus on enhancing its syntax to allow users more flexibility. This includes the ability to define custom actions that better reflect the complexities encountered in real-world situations. Additionally, we plan to incorporate features that let users specify probability distributions for state transitions based on different actions. These enhancements are aimed at making RLML more adaptable and effective for real-world use. By allowing for more detailed and realistic modelling of state transitions and actions, RLML will be better suited to tackle the nuanced and often unpredictable nature of practical RL applications. 
\section{Validation}
\label{sec:validation}

We have validated the proposed reinforcement learning domain-specific language on four applications: path finding, simple game, blackjack, and frozen lake. These applications are well-known applications used in the RL domain \cite{Ravichandiran2018}. 
We present the first three applications here. For details on the use of RLML on the frozen lake application, please refer to \cite{boudakian-mrp-2022}. The artifacts are available at 
    \url{https://github.com/mde-tmu/RLML}.
    
Implementation of the Monte Carlo and DQN algorithms in Java is currently work in progress, hence the validation does not cover these algorithms. 

\subsection{Path Finding Application}
    \begin{figure}[!tbh]
        \centering
        \includegraphics[width=0.55\linewidth]{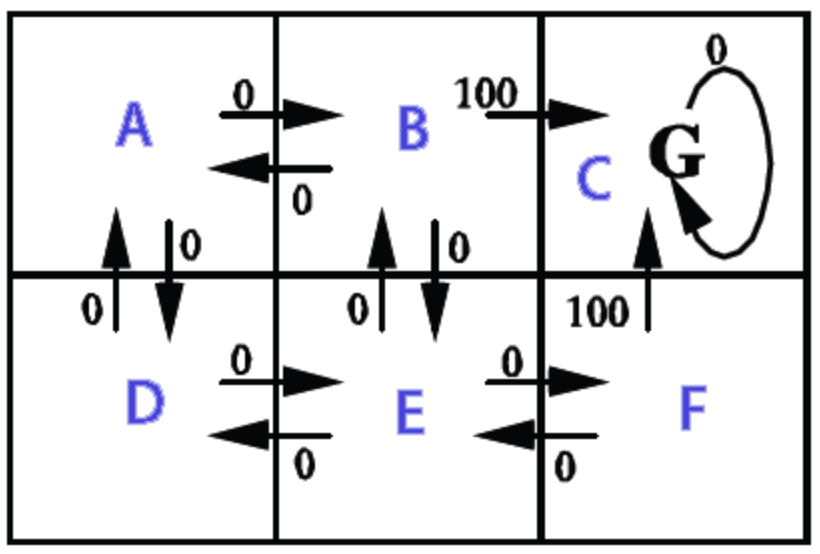}
        \caption{Path Finding Environment}
        \label{fig:PathFindingEnv}
        \end{figure}
        
    The path finding problem \cite{Verma2020} is a common application in the machine learning domain that can be solved with different algorithms, including RL algorithms. In the path finding environment, the agent's goal is to learn the path to a target state, starting from a randomly selected state (see Fig.~\ref{fig:PathFindingEnv}).
    There are in total six states in this application, represented by the alphabetical letters A to F. On each episode the agent will be placed in a random state, from there, it will take actions, and move to new states trying to reach the goal state, which is C for this application. Once the agent reaches the goal state, that episode will be considered complete. The agent will repeat the training episodes a given number of times, as configured in the RL algorithm. In an RLML model, this is set as the total episodes in the RL algorithm entity. At the end of the training, the agent will learn the best path to the goal state C, starting from the random initial state. The agent learns the path to the goal state by updating what is referred to as \emph{Q-Table} and aims to calculate the optimal action value function that can be used to derive the optimal policy. 
    
\begin{figure}
    \centering
    \begin{subfigure}[!tbh]{0.99\textwidth}
        \centering
        \includegraphics[width=1\linewidth]{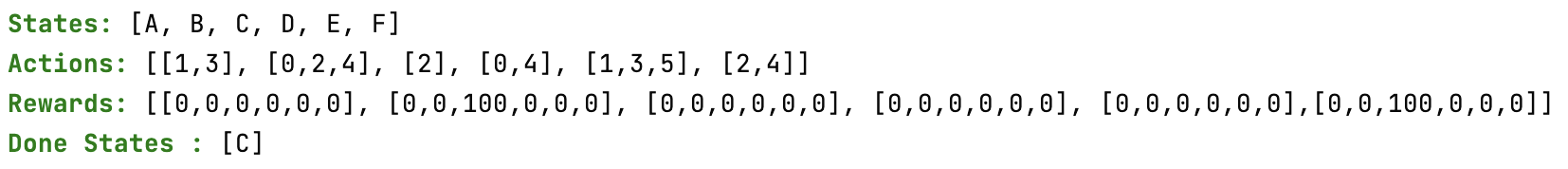}
        \caption{Input Values for Environment}
        \label{fig:PathFindingEnvInput}
    \end{subfigure}
    \begin{subfigure}[!tbh]{0.4\textwidth}
        \centering
        \includegraphics[width=0.85\linewidth]{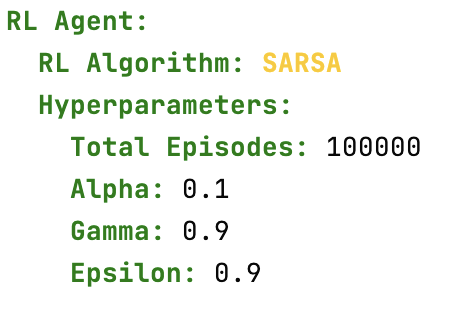}
        \caption{RLML Model with SARSA}
        \label{fig:PathFindingSARSARLML}
    \end{subfigure}
    \begin{subfigure}[!tbh]{0.4\textwidth}
        \centering
        \includegraphics[width=0.85\linewidth]{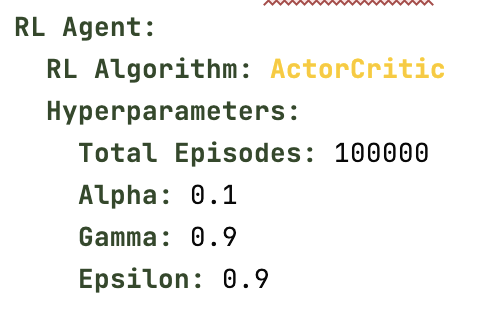}
        \caption{RLML Model with Actor Critic}
        \label{fig:PathFindingACRLML}
    \end{subfigure}
     \caption{Path Finding: RLML Models}
      \label{fig:pathfinding_rlml}
\end{figure}

   The RL environment needs to be modelled in a format that conforms to the RLML abstract and concrete syntax (described in Sec.~\ref{sec:language}). We model the path finding environment as states, actions, rewards and terminal states arrays, as shown in Fig. \ref{fig:PathFindingEnvInput}. Next, the path finding application environment variables and reinforcement learning algorithm option needs to be selected in the RLML model. Sample RLML model instances for the path finding algorithm are shown in Fig.~\ref{fig:PathFindingSARSARLML} and Fig.~\ref{fig:PathFindingACRLML}. 
   
     The general purpose language code is automatically generated from the RLML model for each selected algorithm. 
     At a high level, it is a Java/Python file named according to the RLML root element name property, e.g., \texttt{PathFindingQLearning} (similar to the code presented in Fig.~\ref{fig:SimpleGameSARSAJava} and Fig.~\ref{fig:SimpleGameSARSAPython}). It contains a method to implement the chosen algorithm, a method to run it and one to print the results.

\begin{figure}[!tbh]
        \centering
        \includegraphics[width=0.5\linewidth]{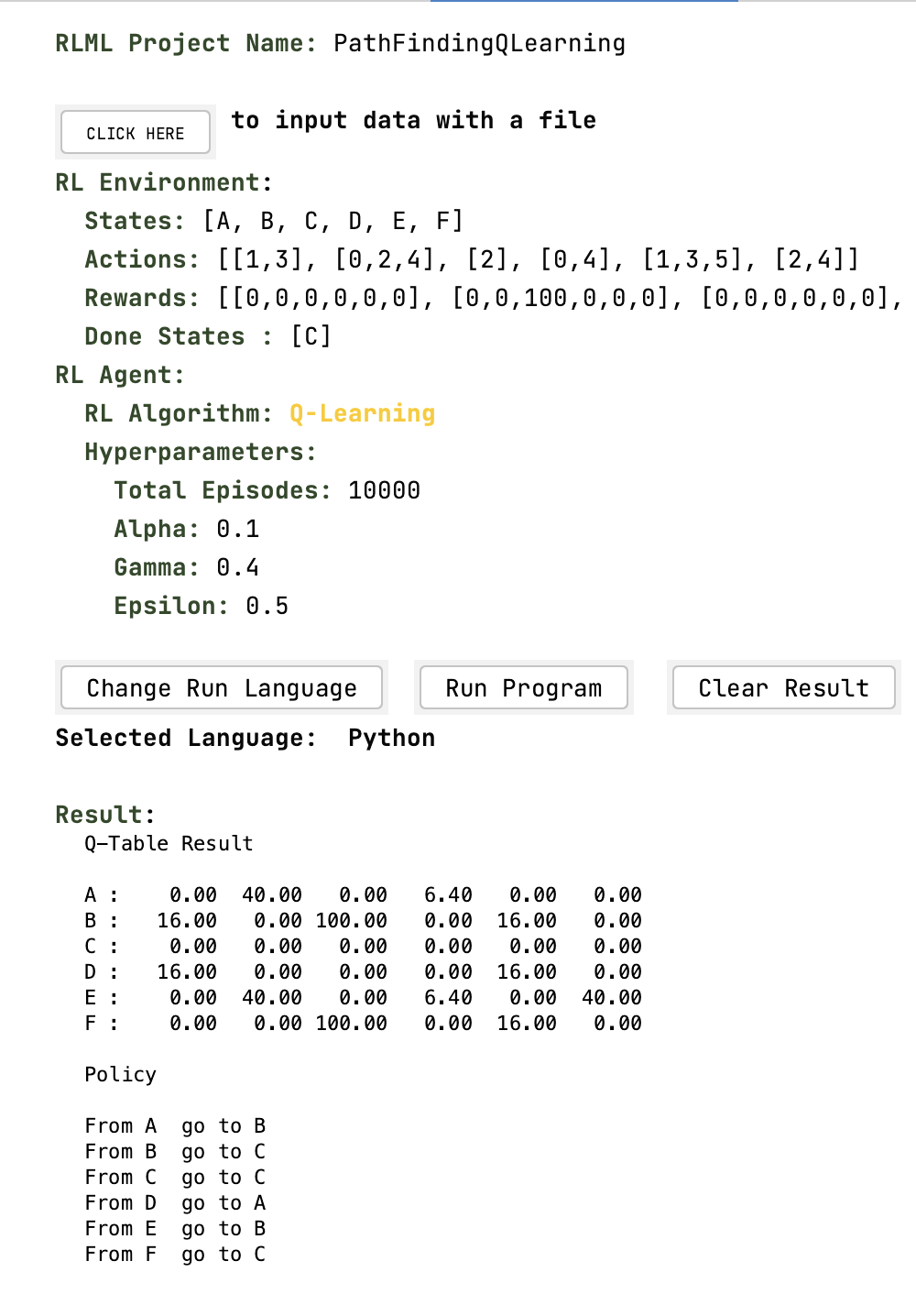}
        \caption{Path Finding Results in RLML with Q-Learning}
        \label{fig:PathFindingQLearningResult}
    \end{figure}

    The RLML modelling environment contains the \texttt{Run Program} button (explained earlier in Sec.~\ref{sec:MPS Environment}). Once we click on the \texttt{Run Program} button, the environment is dynamically updated with the calculated results and we can see the result of running the program in the editor itself. The \emph{Q-Table} and policy are dynamically calculated and displayed. This can be viewed in the modelling environment in the \emph{Results} section (see Fig. \ref{fig:PathFindingQLearningResult}). The policy, derived from \emph{Q-Table} values, shows the preferred action the agent will make at each state. 
    As seen in the results, the agent learned to go to state B from state A, to state C from state B, and so on. Overall, the agent learned the shortest path to go to the target state, which is C.

\begin{figure}[!tbh]
        \centering
        \includegraphics[width=1\linewidth]{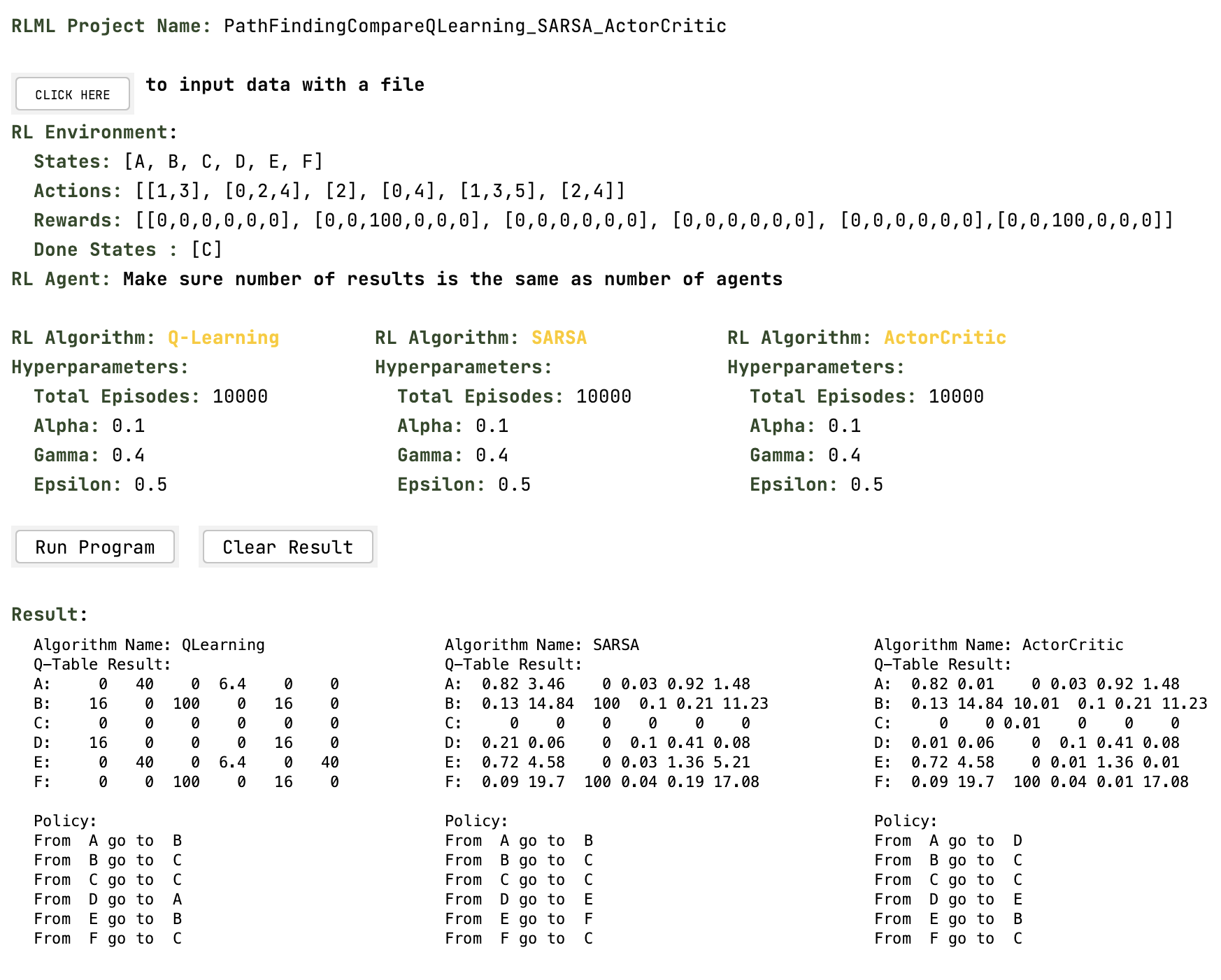}
        \caption{Path Finding Results in RLML: Q-Learning VS SARSA VS ActorCritic}
        \label{fig:PathFindingThreeAlgoResult}
    \end{figure}

    So far the application was implementing the Q-Learning algorithm, however we can easily substitute the algorithm with another algorithm, such as, SARSA (Fig.~\ref{fig:PathFindingSARSARLML}) or Actor Critic algorithm (Fig.~\ref{fig:PathFindingACRLML}). 
    The difference between Q-Learning, SARSA and Actor Critic implementations is minor at the RLML level. RLML only shows the algorithm type and hyperparameters necessary for each algorithm to run. However, it will handle the details of the algorithm calculations during code generation and based on that, it will produce the valid results. As can be seen in Fig.~\ref{fig:PathFindingThreeAlgoResult}, in all cases, the agent successfully learns the path to the goal state.

\subsection{Simple Game Application}
    
\begin{figure}[!tbh]
        \centering
        \includegraphics[width=1\linewidth]{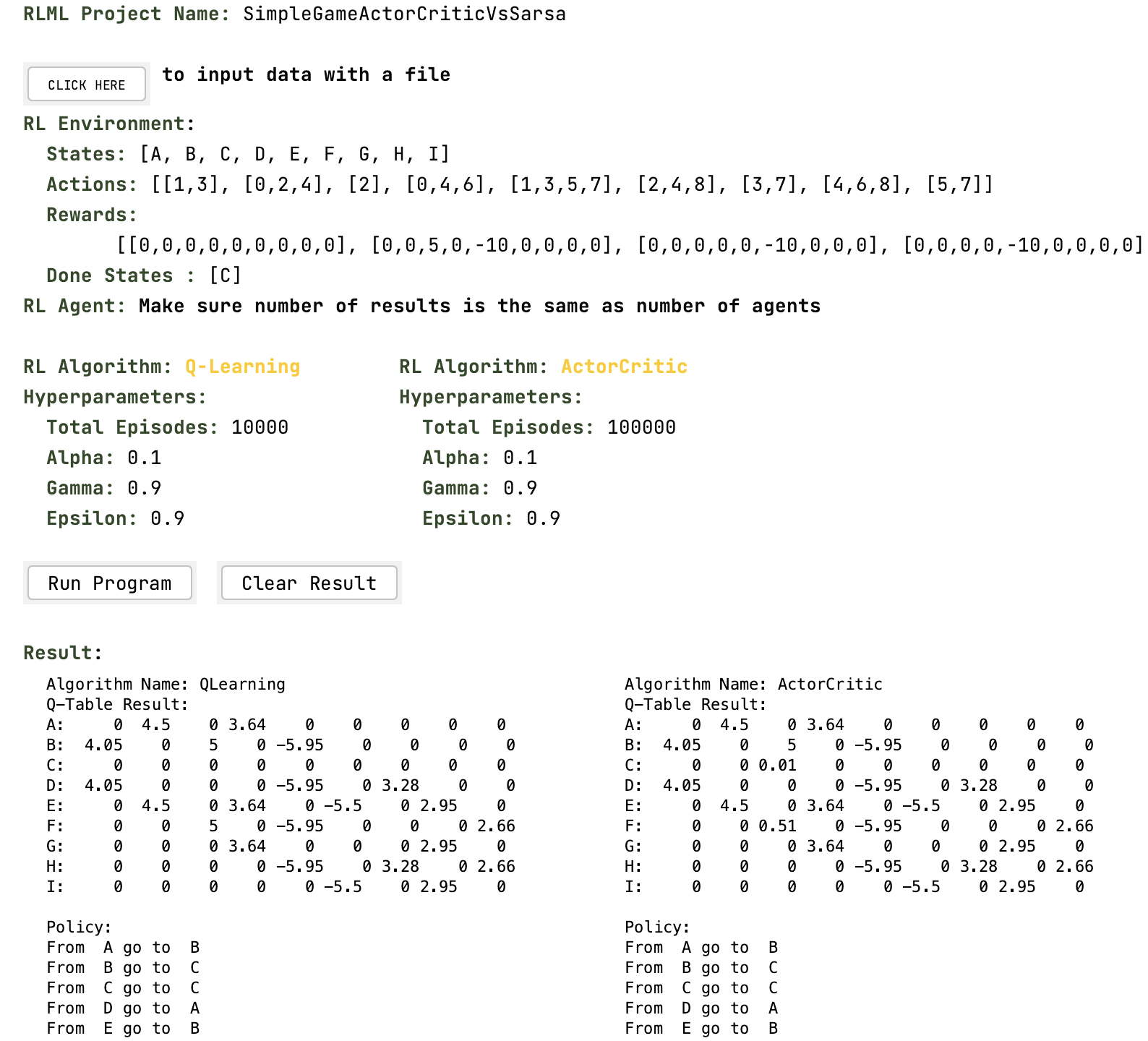}
        \caption{Simple Game Results in RLML: Q-Learning VS ActorCritic}
        \label{fig:SimpleGameResult}
    \end{figure}

    The second example we used to validate RLML is a simple game application \cite{simpleGame} (see Fig. \ref{fig:SimpleGameResult}). This application has a similar environment to the path finding application. The agent here is represented by the player who is targeting a goal state C. In addition to reaching the goal state C, the player needs to avoid danger states which represent some danger concept, for example a fire, on the player in the game environment. The dangerous states reward the player with negative 10. By this negative 10 value, the player will learn to avoid these states and will learn the path to the target state avoiding the dangerous states.
    
    Similar to the path finding application, we need to convert the physical environment of this simple game application to values RLML can work with. The input values for the simple game RL problem environment are shown in Fig.~\ref{fig:SimpleGameResult}. The RLML models (only Q-learning and Actor Critic included for space reasons) are shown here.

\begin{figure}[!tbh]
    \centering
    \includegraphics[width=\linewidth]{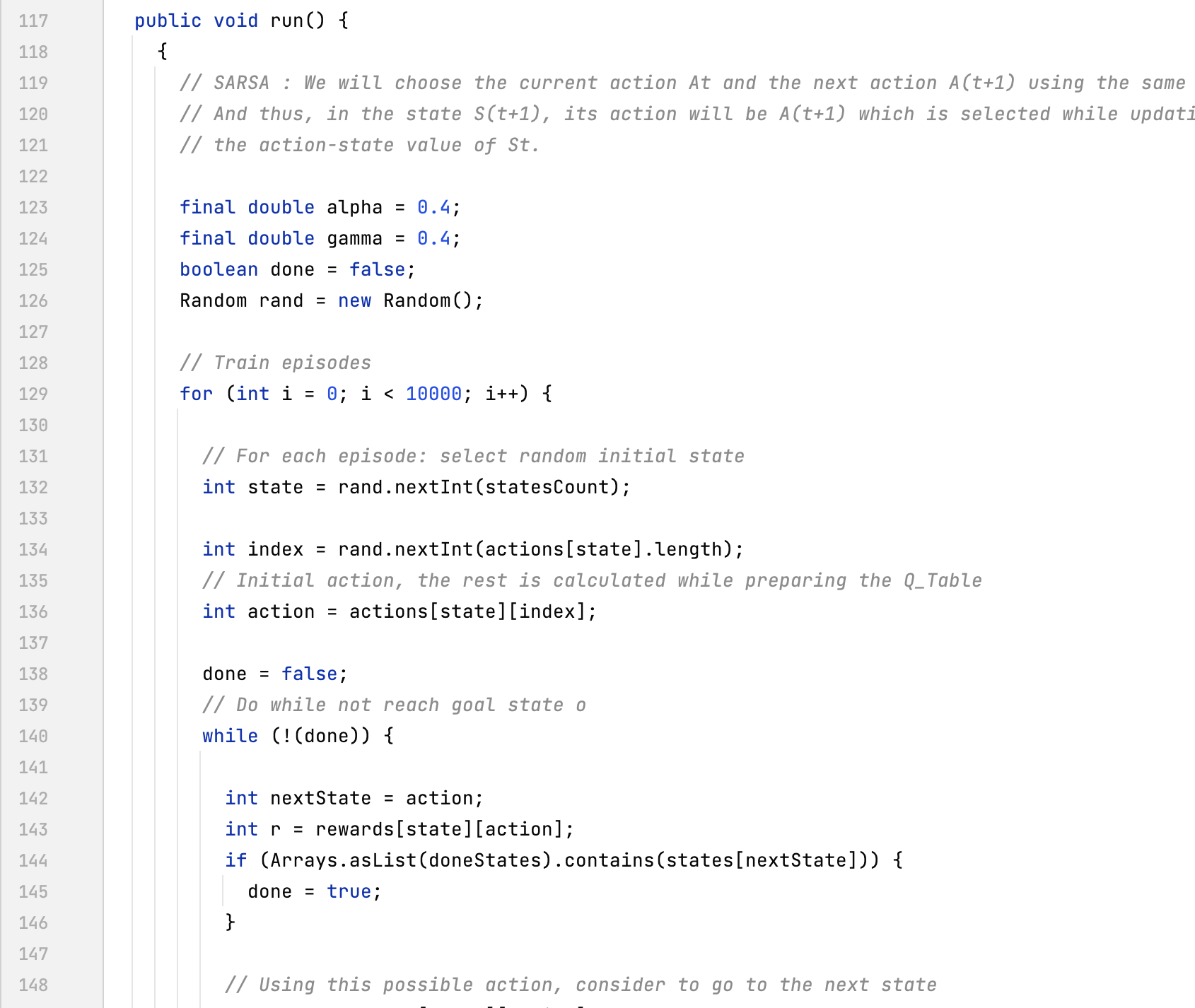}
    \caption{Simple Game with SARSA: Generated Java Code with RLML}
    \label{fig:SimpleGameSARSAJava}
\end{figure}

\begin{figure}[!tbh]
    \centering
    \includegraphics[width=\linewidth]{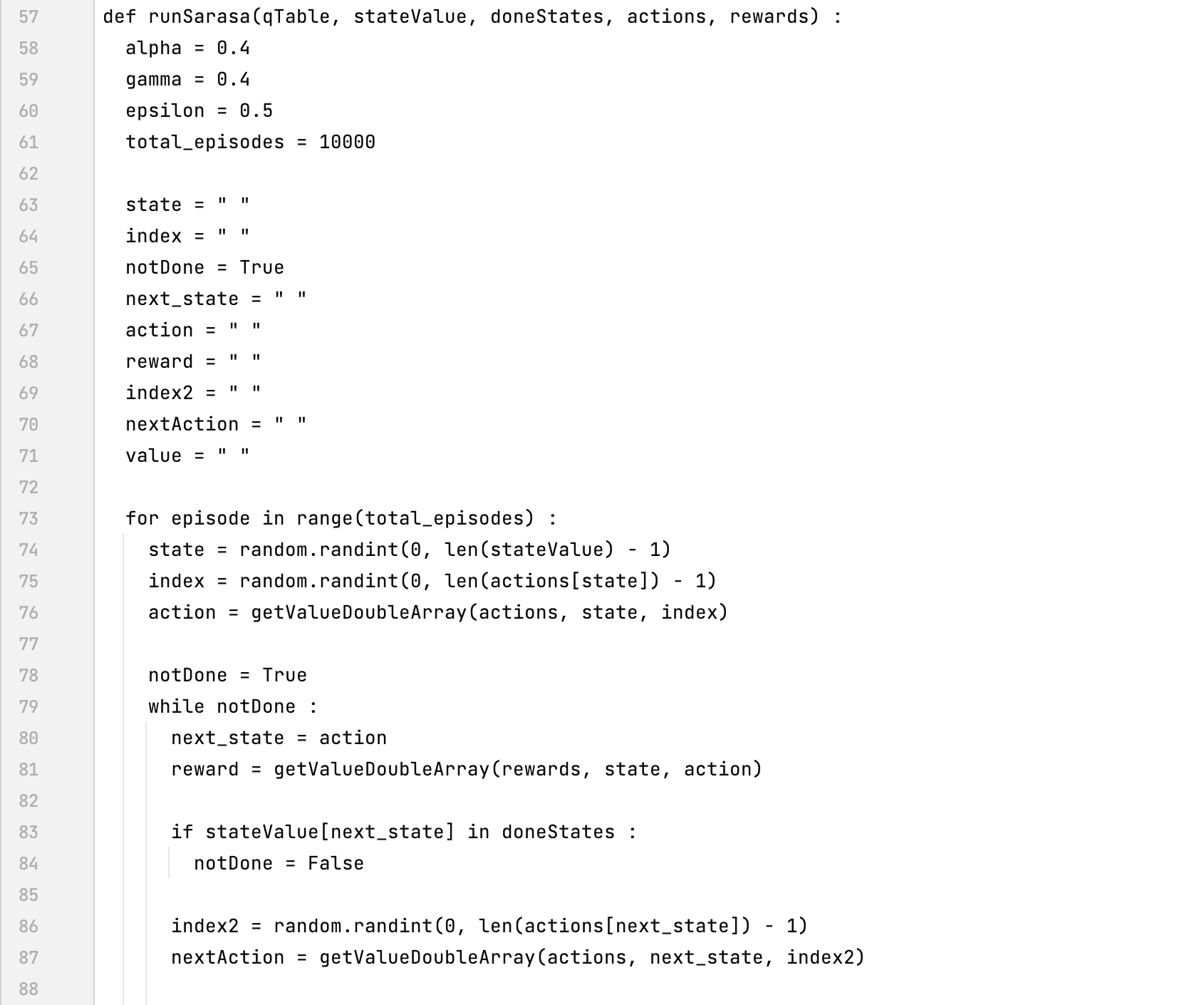}
    \caption{Simple Game with SARSA: Generated Python Code with RLML}
    \label{fig:SimpleGameSARSAPython}
\end{figure}

    As we saw earlier in the path finding application, we can get the generated Java code as a class (e.g., named \texttt{SimpleGameQLearning} as presented in Fig.~\ref{fig:SimpleGameSARSAJava}) or the generated Python code (as presented in Fig.~\ref{fig:SimpleGameSARSAPython}), run the program and get the results with the click of a button ``Run Program" within the RLML editor.
    As an example, if we look at the calculated \emph{Q-Table} and policy in Fig.~\ref{fig:SimpleGameResult}, 
    we can notice that the player learned the path to the goal state C, and also learned to avoid the danger states E and F, because in the \emph{Q-Table} values, states F and E are associated with negative values. Following this policy, the agent will avoid the states with negative values. In the printed policy path, we can see that the player never takes E or F states options. For example, from H it goes to G, and from I it goes to H, and not F. Even though there are fewer steps to reach to the goal state C through the danger states, the player will not take that path, because the player or the agent has learned to avoid the path with danger states, since it receives negative rewards.

\subsection{Blackjack Game}
\begin{figure}[!tbh]
        \centering
        \includegraphics[width=1\linewidth]{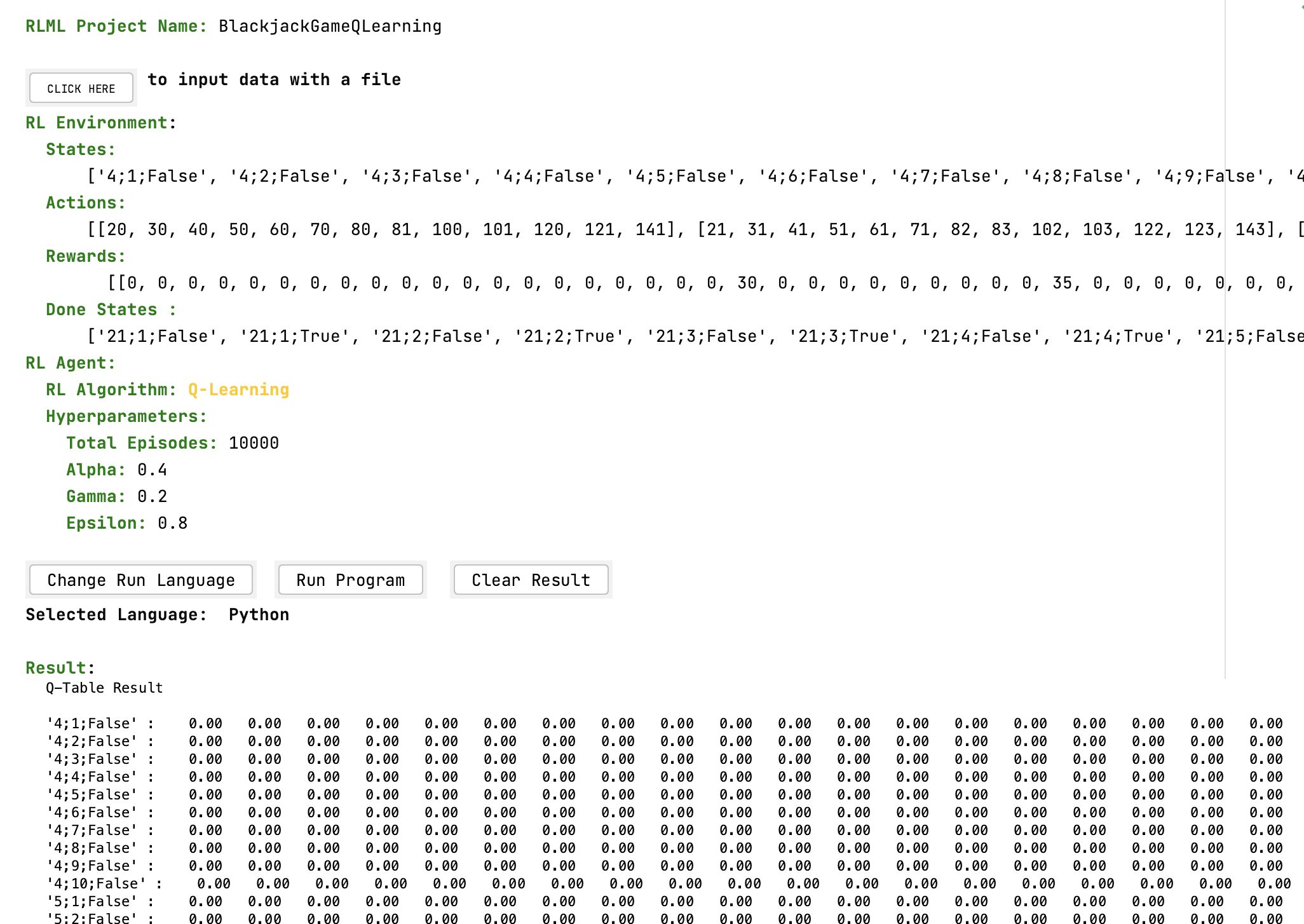}
        \caption{Blackjack Results in RLML with Q-Learning}
        \label{fig:BlackjackResultSnippet}
    \end{figure}

The Blackjack game is a prominent application in the domain of machine learning, solvable through various algorithms, including those from reinforcement learning (RL). In the Blackjack environment, the objective of the agent is to master decision-making strategies to maximize winnings, beginning from an initial hand (see Fig.~\ref{fig:BlackjackResultSnippet}). This involves understanding when to hit, stand, or make other game-specific decisions based on the current hand and the dealer's visible card, aiming to attain a hand value as close to 21 as possible without exceeding it. Given the concrete syntax of the language going to any state but the current state counts as \emph{``Hit"} action, whereas staying at current state counts as \emph{``Stand"}

Since, the game represents a real world application the state space is quite complex consisting of around 460 states. In each episode, the agent starts with an initial hand in Blackjack, taking actions based on the hand's value and the dealer's visible card, with the aim of optimizing its strategy for winning. The goal in this context is to make decisions that maximize the agent's chances of beating the dealer without exceeding a hand value of 21. Each completed hand is an episode, and the agent undergoes numerous episodes as defined in the RL algorithm's settings. In the RLML model, this is specified as the total number of episodes under the RL algorithm entity. Through training, the agent learns the best decision-making strategy for Blackjack, starting from any given hand. It achieves this by updating the Q-Table, with the ultimate aim of determining the optimal action-value function to derive the optimal policy.
Similar to the pathfinding application, the RL environment needs to be modelled in a format that conforms to the RLML abstract and concrete syntax.
In line with that, we can generate and run code for a Blackjack game (e.g., named \emph{BlackjackQLearning}) using the RLML editor's ``Run Program" button. Examining the calculated Q-Table and derived policy reveals how the player has learned optimal decision-making strategies in the Blackjack environment. In the Q-Table, specific actions in certain states may have negative values, reflecting decisions that typically lead to losing hands.

The algorithm learns to avoid actions that historically result in losses, even if they seem initially appealing. The player, or agent, has been trained to prioritize decisions that maximize overall gains over time, rather than immediate, riskier gains, as indicated by the positive and negative rewards in the Q-Table.   

    \begin{figure}[!tbh]
        \centering
        \includegraphics[width=1\linewidth]{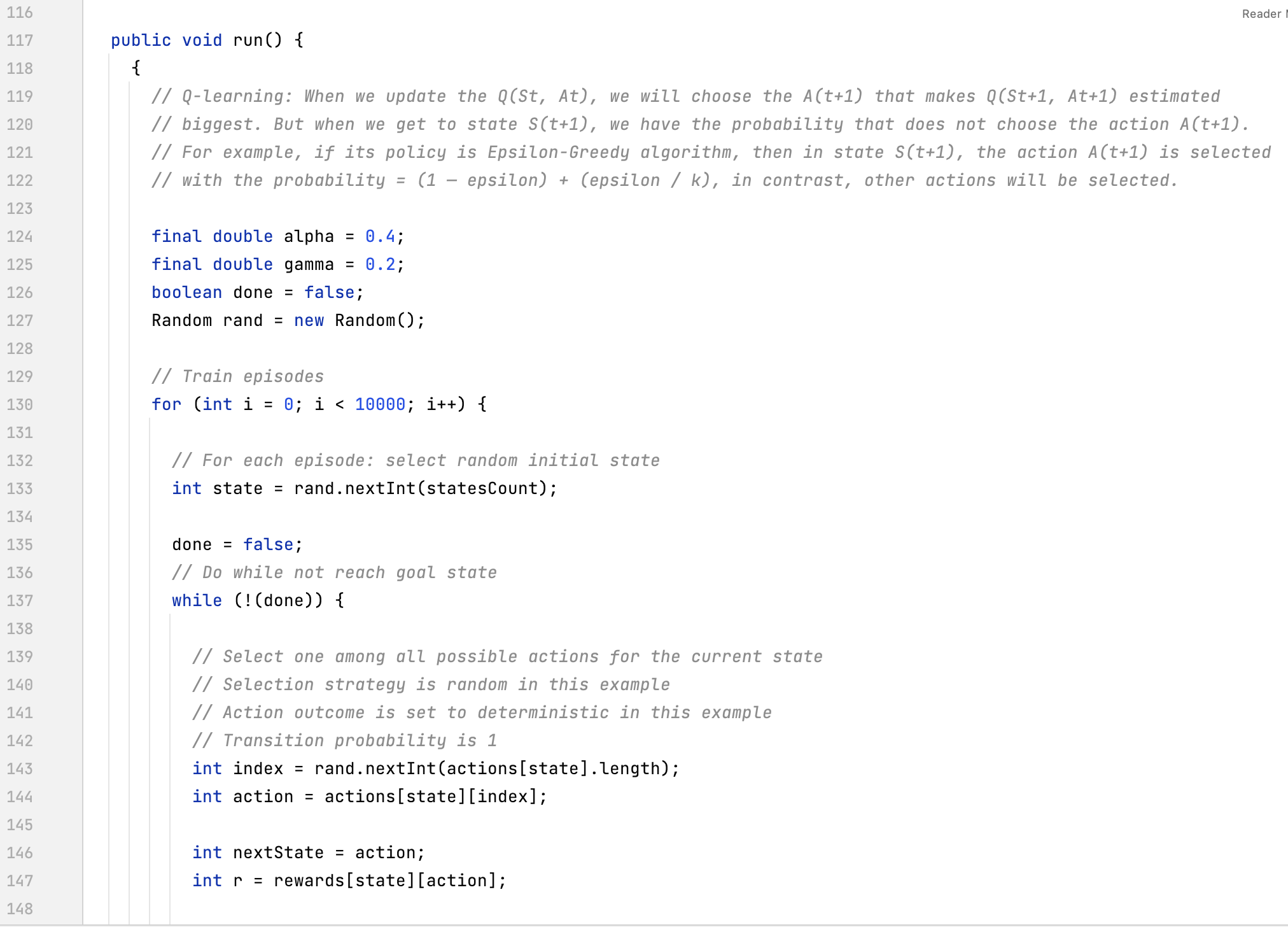}
        \caption{Blackjack with Q-Learning: Generated Code with RLML}
        \label{fig:BlackjackResultCodeGen}
    \end{figure}

\section{Related Work}
\label{sec:relatedwork}

While machine learning is widely applied in the MDE area, there is a need for MDE application in the ML area~\cite{Bucchiarone2020}. There is limited work available in this area. DSMLs for the artificial intelligence domain and more specifically, machine learning domain, are recent contributions, driven by the need to make ML algorithms easier to understand by improving the learning curve, as well as more attractive to apply and faster to test and compare different solutions to one another. The application of MDE in the RL domain, in particular, is sparse. To the best of our knowledge, this is the first work on developing a domain-specific language for reinforcement learning.

In this section, we discuss some relevant ML/RL projects and papers that involve some form of MDE. We also present an overview of existing RL libraries and toolkits.

\subsection{Application of MDE in ML}
    
    Our work takes inspiration from the Classification Algorithm Framework (CAF) \cite{meacham2020}. The primary difference between CAF and RLML is that CAF is developed for machine learning classification algorithms, and RLML is developed for reinforcement learning algorithms. CAF expects inputs to apply classification algorithms, and RLML expects inputs to implement reinforcement learning algorithm. They have similar configuration-like interface where user can interact with it. CAF supports code generation in the Java language, while RLML supports both Java and Python. Unlike CAF, RLML offers a comparator feature.

    Liaskos et al.~\cite{liaskos2022} present a modelling and design process for generating simulation environments for RL based on goal models defined using iStar. This allows model-based reasoning to be carried out and for agents to be trained prior to deployment in the target environment. 
    The paper mentions that high-level RL models can be automatically mapped to these simulation components. While this work has a very different goal than RLML, it can be  complementary to RLML, such that RLML models can be translated to goal models to provide simulation support. 
    
    ThingML2\cite{moin2020_thingml2} is preliminary work on a modelling language for the Internet of Things (IoT) domain integrated with ML algorithms. Their goal is to connect software models with ML models to produce or interact with data analytics models.
    
    DeepDSL \cite{Zhao2017} is a domain-specific language embedded in Scala, for developing deep learning applications. 
    It provides compiler-based optimizations for deep learning models to run with less memory usage and/or in shorter time. DeepDSL allows users to define deep learning networks as tensor functions 
    and has its own compiler that produces DeepDSL Java program. Therefore, it is quite different from RLML, which is developed with the MPS language workbench for the RL domain.
    
    The literature overview varies in technologies and approaches. Our proposed approach is unique and different than the rest of the reviewed work, because we use model-driven engineering to create abstractions in the RL domain. 

\subsection{RL Libraries and Toolkits}

As the field of reinforcement learning evolves, various platforms and libraries have been established to facilitate the development of RL applications, each with its distinct features and focus areas. For instance, the Reinforcement Learning Toolkit developed with Unreal Engine \cite{sapio-hcii2022} emphasizes immersive simulation environments. Python libraries such as RL-coach \cite{RL-coach}, Tensorforce \cite{tensorflow2015-whitepaper}, TRFL \cite{TRFL}, and TF Agents \cite{TFAgents} offer robust support for RL algorithms but require developers to engage deeply with algorithmic details.

These existing tools, written in Python, demand substantial technical understanding of RL processes, from algorithm implementation to the handling of complex calculations. This requirement can be a barrier for those who are not specialists in RL as well as for non-programmers.

In this context, the development of RLML, as discussed in the paper, represents a novel approach within the RL landscape. By focusing on this higher level of abstraction, RLML seeks to simplify the use of RL with a user-friendly modelling environment, reducing the necessity for in-depth knowledge of the underlying algorithms.

A recent addition to these set of tools is Scikit-decide \cite{ScikitDecide}, an AI framework for RL, automated planning and scheduling. It enhances the Scikit ecosystem by simplifying RL algorithm implementations, paralleling Scikit-learn's approach in classical machine learning. While it offers an intuitive interface, understanding of RL fundamentals is still essential, a gap RLML seeks to bridge with its user-friendly domain-specific modelling language.

This contribution could be particularly valuable in making advanced RL technologies more approachable for non-data scientists or for users without extensive background in the field, thereby expanding the reach and application of RL in various domains.
\section{Conclusion}
\label{sec:conclusion}

We proposed a domain-specific language, RLML, for the reinforcement learning domain. 
With the use of the language workbench MPS, we built a domain-specific modelling environment for RLML, which supports model editing, syntax checking, constraints checking and validation, as well as code generation. Our proposed language is developed to be easily extensible to support a wide range of RL algorithms. However in this paper, we targeted model-free, gradient-free category of reinforcement learning algorithms. To the best of our knowledge, this work is a first step in this direction for reinforcement learning. 
    
Through our reinforcement learning applications, we validated our proposed language. With our validation, we showcased how RLML achieved the abstraction needed in RL applications, by providing a configuration-like model where it is only expecting input values of the RL problem environment and a choice of RL algorithm. From that point, the RLML modelling environment can generate executable code, run it and display the results. The environment also provides a comparator to compare results obtained with different RL algorithms. It supports both Java and Python implementations.  

RLML can also be used on business case studies and to get feedback from reinforcement learning users at different levels of expertise. Moreover, RLML can be helpful in academia for making reinforcement learning accessible for non-technical students.
This work is a starting point towards developing an environment for supporting various types of RL technologies, both model-free and model-based. As discussed earlier, we are working on incorporating probability distributions and custom actions into RLML which will allow it to model real world use cases more effectively.

\backmatter

\bmhead{Acknowledgements}

This work has been partially supported by Natural Sciences and Engineering Research Council of Canada (NSERC) and Toronto Metropolitan University. The authors would like to extend their thanks to Prof. Nariman Farsad for his feedback on this work.


\bibliographystyle{sn-mathphy-num}
\bibliography{bibliography}


\begin{thebibliography}{26}
\ifx \bisbn   \undefined \def \bisbn  #1{ISBN #1}\fi
\ifx \binits  \undefined \def \binits#1{#1}\fi
\ifx \bauthor  \undefined \def \bauthor#1{#1}\fi
\ifx \batitle  \undefined \def \batitle#1{#1}\fi
\ifx \bjtitle  \undefined \def \bjtitle#1{#1}\fi
\ifx \bvolume  \undefined \def \bvolume#1{\textbf{#1}}\fi
\ifx \byear  \undefined \def \byear#1{#1}\fi
\ifx \bissue  \undefined \def \bissue#1{#1}\fi
\ifx \bfpage  \undefined \def \bfpage#1{#1}\fi
\ifx \blpage  \undefined \def \blpage #1{#1}\fi
\ifx \burl  \undefined \def \burl#1{\textsf{#1}}\fi
\ifx \doiurl  \undefined \def \doiurl#1{\url{https://doi.org/#1}}\fi
\ifx \betal  \undefined \def \betal{\textit{et al.}}\fi
\ifx \binstitute  \undefined \def \binstitute#1{#1}\fi
\ifx \binstitutionaled  \undefined \def \binstitutionaled#1{#1}\fi
\ifx \bctitle  \undefined \def \bctitle#1{#1}\fi
\ifx \beditor  \undefined \def \beditor#1{#1}\fi
\ifx \bpublisher  \undefined \def \bpublisher#1{#1}\fi
\ifx \bbtitle  \undefined \def \bbtitle#1{#1}\fi
\ifx \bedition  \undefined \def \bedition#1{#1}\fi
\ifx \bseriesno  \undefined \def \bseriesno#1{#1}\fi
\ifx \blocation  \undefined \def \blocation#1{#1}\fi
\ifx \bsertitle  \undefined \def \bsertitle#1{#1}\fi
\ifx \bsnm \undefined \def \bsnm#1{#1}\fi
\ifx \bsuffix \undefined \def \bsuffix#1{#1}\fi
\ifx \bparticle \undefined \def \bparticle#1{#1}\fi
\ifx \barticle \undefined \def \barticle#1{#1}\fi
\bibcommenthead
\ifx \bconfdate \undefined \def \bconfdate #1{#1}\fi
\ifx \botherref \undefined \def \botherref #1{#1}\fi
\ifx \url \undefined \def \url#1{\textsf{#1}}\fi
\ifx \bchapter \undefined \def \bchapter#1{#1}\fi
\ifx \bbook \undefined \def \bbook#1{#1}\fi
\ifx \bcomment \undefined \def \bcomment#1{#1}\fi
\ifx \oauthor \undefined \def \oauthor#1{#1}\fi
\ifx \citeauthoryear \undefined \def \citeauthoryear#1{#1}\fi
\ifx \endbibitem  \undefined \def \endbibitem {}\fi
\ifx \bconflocation  \undefined \def \bconflocation#1{#1}\fi
\ifx \arxivurl  \undefined \def \arxivurl#1{\textsf{#1}}\fi
\csname PreBibitemsHook\endcsname

\bibitem[\protect\citeauthoryear{Brunton and Kutz}{2019}]{Brunton2019}
\begin{bbook}
\bauthor{\bsnm{Brunton}, \binits{S.}},
\bauthor{\bsnm{Kutz}, \binits{J.}}:
\bbtitle{Data-Driven Science and Engineering}.
\bpublisher{Cambridge University Press},
\blocation{Cambridge}
(\byear{2019}).
\doiurl{10.1017/9781108380690.001}
\end{bbook}
\endbibitem

\bibitem[\protect\citeauthoryear{Bucchiarone et~al.}{2020}]{Bucchiarone2020}
\begin{barticle}
\bauthor{\bsnm{Bucchiarone}, \binits{A.}},
\bauthor{\bsnm{Cabot}, \binits{J.}},
\bauthor{\bsnm{Paige}, \binits{R.}},
\bauthor{\bsnm{Pierantonio}, \binits{A.}}:
\batitle{Grand challenges in model-driven engineering: an analysis of the state of the research}.
\bjtitle{Software and Systems Modeling}
\bvolume{19},
\bfpage{1}--\blpage{9}
(\byear{2020})
\doiurl{10.1007/s10270-019-00773-6}
\end{barticle}
\endbibitem

\bibitem[\protect\citeauthoryear{Baier et~al.}{2019}]{Baier2019}
\begin{bchapter}
\bauthor{\bsnm{Baier}, \binits{L.}},
\bauthor{\bsnm{Jöhren}, \binits{F.}},
\bauthor{\bsnm{Seebacher}, \binits{S.}}:
\bctitle{Challenges in the deployment and operation of machine learning in practice}.
In: \bbtitle{27th European Conference on Information Systems (ECIS 2019)},
vol. \bseriesno{1}
(\byear{2019})
\end{bchapter}
\endbibitem

\bibitem[\protect\citeauthoryear{Voelter et~al.}{2013}]{Voelter2013}
\begin{bbook}
\bauthor{\bsnm{Voelter}, \binits{M.}},
\bauthor{\bsnm{Benz}, \binits{S.}},
\bauthor{\bsnm{Dietrich}, \binits{C.}},
\bauthor{\bsnm{Engelmann}, \binits{B.}},
\bauthor{\bsnm{Helander}, \binits{M.}},
\bauthor{\bsnm{Kats}, \binits{L.C.L.}},
\bauthor{\bsnm{Visser}, \binits{E.}},
\bauthor{\bsnm{Wachsmuth}, \binits{G.}}:
\bbtitle{DSL Engineering - Designing, Implementing and Using Domain-Specific Languages},
pp. \bfpage{1}--\blpage{558}.
\bpublisher{CreateSpace Independent Publishing Platform},
\blocation{US}
(\byear{2013})
\end{bbook}
\endbibitem

\bibitem[\protect\citeauthoryear{Schmidt}{2006}]{Schmidt2006}
\begin{barticle}
\bauthor{\bsnm{Schmidt}, \binits{D.C.}}:
\batitle{Model-driven engineering}.
\bjtitle{Computer-IEEE Computer Society}
\bvolume{39}(\bissue{2}),
\bfpage{25}
(\byear{2006})
\end{barticle}
\endbibitem

\bibitem[\protect\citeauthoryear{Graesser and Keng}{2019}]{Graesser2019}
\begin{bbook}
\bauthor{\bsnm{Graesser}, \binits{L.}},
\bauthor{\bsnm{Keng}, \binits{W.L.}}:
\bbtitle{Foundations of Deep Reinforcement Learning: Theory and Practice in Python}.
\bpublisher{Addison-Wesley Professional},
\blocation{Boston}
(\byear{2019})
\end{bbook}
\endbibitem

\bibitem[\protect\citeauthoryear{Harel}{1987}]{harel1987statecharts}
\begin{barticle}
\bauthor{\bsnm{Harel}, \binits{D.}}:
\batitle{Statecharts: A visual formalism for complex systems}.
\bjtitle{Science of computer programming}
\bvolume{8}(\bissue{3}),
\bfpage{231}--\blpage{274}
(\byear{1987})
\end{barticle}
\endbibitem

\bibitem[\protect\citeauthoryear{Sutton and Barto}{1999}]{Montague1999}
\begin{barticle}
\bauthor{\bsnm{Sutton}, \binits{R.S.}},
\bauthor{\bsnm{Barto}, \binits{A.G.}}:
\batitle{Reinforcement learning: An introduction}.
\bjtitle{Trends in cognitive sciences}
\bvolume{3}(\bissue{9}),
\bfpage{360}--\blpage{360}
(\byear{1999})
\end{barticle}
\endbibitem

\bibitem[\protect\citeauthoryear{Silver}{2015}]{silver2015}
\begin{botherref}
\oauthor{\bsnm{Silver}, \binits{D.}}:
Lectures on reinforcement learning
(2015)
\end{botherref}
\endbibitem

\bibitem[\protect\citeauthoryear{Jiang}{2021}]{Jiang2021}
\begin{bbook}
\bauthor{\bsnm{Jiang}, \binits{H.}}:
\bbtitle{Machine Learning Fundamentals: a Concise Introduction}.
\bpublisher{Cambridge University Press},
\blocation{Cambridge}
(\byear{2021})
\end{bbook}
\endbibitem

\bibitem[\protect\citeauthoryear{}{}]{MPS}
\begin{botherref}
{MPS}: The Domain-Specific Language Creator by JetBrains.
\url{https://www.jetbrains.com/mps/}
\end{botherref}
\endbibitem

\bibitem[\protect\citeauthoryear{Radford et~al.}{2019}]{radford2019-openai}
\begin{botherref}
\oauthor{\bsnm{Radford}, \binits{A.}},
\oauthor{\bsnm{Wu}, \binits{J.}},
\oauthor{\bsnm{Amodei}, \binits{D.}},
\oauthor{\bsnm{Amodei}, \binits{D.}},
\oauthor{\bsnm{Clark}, \binits{J.}},
\oauthor{\bsnm{Brundage}, \binits{M.}},
\oauthor{\bsnm{Sutskever}, \binits{I.}}:
Better language models and their implications.
OpenAI blog
\textbf{1}(2)
(2019)
\end{botherref}
\endbibitem

\bibitem[\protect\citeauthoryear{Ravichandiran}{2018}]{Ravichandiran2018}
\begin{bbook}
\bauthor{\bsnm{Ravichandiran}, \binits{S.}}:
\bbtitle{Hands-On Reinforcement Learning with Python: Master Reinforcement and Deep Reinforcement Learning Using {OpenAI} Gym and TensorFlow}.
\bpublisher{Packt Publishing, Limited},
\blocation{Birmingham}
(\byear{2018})
\end{bbook}
\endbibitem

\bibitem[\protect\citeauthoryear{Sinani}{2022}]{boudakian-mrp-2022}
\begin{botherref}
\oauthor{\bsnm{Sinani}, \binits{N.}}:
{RLML}: A Domain-Specific Modelling Language for Reinforcement Learning,
Toronto, ON.
{MRP} Report
(2022)
\end{botherref}
\endbibitem

\bibitem[\protect\citeauthoryear{Verma et~al.}{2020}]{Verma2020}
\begin{barticle}
\bauthor{\bsnm{Verma}, \binits{P.}},
\bauthor{\bsnm{Dhanre}, \binits{U.}},
\bauthor{\bsnm{Khekare}, \binits{S.}},
\bauthor{\bsnm{Sheikh}, \binits{S.}},
\bauthor{\bsnm{Khekare}, \binits{G.}}:
\batitle{The optimal path finding algorithm based on reinforcement learning}.
\bjtitle{International Journal of Software Science and Computational Intelligence}
\bvolume{12},
\bfpage{1}--\blpage{18}
(\byear{2020})
\doiurl{10.4018/IJSSCI.2020100101}
\end{barticle}
\endbibitem

\bibitem[\protect\citeauthoryear{Doshi}{2020}]{simpleGame}
\begin{botherref}
\oauthor{\bsnm{Doshi}, \binits{K.}}:
Reinforcement Learning Explained Visually (Part 4): Q Learning, step-by-step
(2020).
\url{https://towardsdatascience.com/reinforcement-learning-explained-visually-part-4-q-learning-step-by-step-b65efb731d3e}
\end{botherref}
\endbibitem

\bibitem[\protect\citeauthoryear{Meacham et~al.}{2020}]{meacham2020}
\begin{barticle}
\bauthor{\bsnm{Meacham}, \binits{S.}},
\bauthor{\bsnm{Pech}, \binits{V.}},
\bauthor{\bsnm{Nauck}, \binits{D.}}:
\batitle{Classification algorithms framework {(CAF)} to enable intelligent systems using jetbrains {MPS} domain-specific languages environment}.
\bjtitle{IEEE access}
\bvolume{8},
\bfpage{14832}--\blpage{14840}
(\byear{2020})
\end{barticle}
\endbibitem

\bibitem[\protect\citeauthoryear{Liaskos et~al.}{2022}]{liaskos2022}
\begin{bchapter}
\bauthor{\bsnm{Liaskos}, \binits{S.}},
\bauthor{\bsnm{Khan}, \binits{S.M.}},
\bauthor{\bsnm{Golipour}, \binits{R.}},
\bauthor{\bsnm{Mylopoulos}, \binits{J.}}:
\bctitle{Towards goal-based generation of reinforcement learning domain simulations.}
In: \bbtitle{iStar},
pp. \bfpage{22}--\blpage{28}
(\byear{2022})
\end{bchapter}
\endbibitem

\bibitem[\protect\citeauthoryear{Moin et~al.}{2020}]{moin2020_thingml2}
\begin{bchapter}
\bauthor{\bsnm{Moin}, \binits{A.}},
\bauthor{\bsnm{R\"{o}ssler}, \binits{S.}},
\bauthor{\bsnm{Sayih}, \binits{M.}},
\bauthor{\bsnm{G\"{u}nnemann}, \binits{S.}}:
\bctitle{From things' modeling language ({ThingML}) to things' machine learning ({ThingML2})}.
In: \bbtitle{Proceedings of the 23rd ACM/IEEE International Conference on Model Driven Engineering Languages and Systems: Companion Proceedings (MODELS-C)},
pp. \bfpage{1}--\blpage{2}.
\bpublisher{ACM}, \blocation{???}
(\byear{2020}).
\doiurl{10.1145/3417990.3420057}
\end{bchapter}
\endbibitem

\bibitem[\protect\citeauthoryear{Zhao et~al.}{2017}]{Zhao2017}
\begin{botherref}
\oauthor{\bsnm{Zhao}, \binits{T.}},
\oauthor{\bsnm{Huang}, \binits{X.}},
\oauthor{\bsnm{Cao}, \binits{Y.}}:
{DeepDSL}: A compilation-based domain-specific language for deep learning
(2017)
\end{botherref}
\endbibitem

\bibitem[\protect\citeauthoryear{Sapio and Ratini}{2022}]{sapio-hcii2022}
\begin{bchapter}
\bauthor{\bsnm{Sapio}, \binits{F.}},
\bauthor{\bsnm{Ratini}, \binits{R.}}:
\bctitle{Developing and testing a new reinforcement learning toolkit with unreal engine}.
In: \bbtitle{Artificial Intelligence in HCI},
pp. \bfpage{317}--\blpage{334}.
\bpublisher{Springer},
\blocation{Cham}
(\byear{2022})
\end{bchapter}
\endbibitem

\bibitem[\protect\citeauthoryear{}{}]{RL-coach}
\begin{botherref}
Reinforcement Learning Coach {RL}-coach.
\url{https://intellabs.github.io/coach/}
\end{botherref}
\endbibitem

\bibitem[\protect\citeauthoryear{Abadi et~al.}{2016}]{tensorflow2015-whitepaper}
\begin{botherref}
\oauthor{\bsnm{Abadi}, \binits{M.}},
\oauthor{\bsnm{Agarwal}, \binits{A.}},
\oauthor{\bsnm{Barham}, \binits{P.}},
\oauthor{\bsnm{Brevdo}, \binits{E.}},
\oauthor{\bsnm{al.}}:
{TensorFlow}: Large-scale machine learning on heterogeneous systems.
CoRR
\textbf{abs/1603.04467}
(2016)
\end{botherref}
\endbibitem

\bibitem[\protect\citeauthoryear{}{}]{TRFL}
\begin{botherref}
{Deepmind TRFL}.
\url{https://www.deepmind.com/open-source/trfl}
\end{botherref}
\endbibitem

\bibitem[\protect\citeauthoryear{}{}]{TFAgents}
\begin{botherref}
TensorFlow Agents ({TF} Agents).
\url{https://www.tensorflow.org/agents}
\end{botherref}
\endbibitem

\bibitem[\protect\citeauthoryear{}{}]{ScikitDecide}
\begin{botherref}
Scikit Decide.
\url{https://airbus.github.io/scikit-decide/guide/introduction}
\end{botherref}
\endbibitem

\end{thebibliography}

\end{document}